\documentclass[twoside,
 twocolumn,
 superscriptaddress]{revtex4-2}
 
\usepackage{hyperref}
\hypersetup{colorlinks=true,breaklinks=true, urlcolor=blue,linkcolor=blue,citecolor=red}
\usepackage{times}
\usepackage{graphicx}
\usepackage{dcolumn}
\usepackage{bm}
\usepackage{amsmath}

\usepackage{enumitem}

\makeatletter

\newcommand{\mat}[1]{{\mathpalette\mat@{#1}}}
\newcommand{\mat@}[2]{%
 \begingroup
 \sbox\z@{$\m@th#1\underline{#2}$}%
 \dimen@=\dp\z@ \advance\dimen@ -2\mat@dimen{#1}%
 \dp\z@=\dimen@
 \sbox\z@{$\m@th\underline{\box\z@}$}%
 \box\z@
 \endgroup
}
\newcommand\mat@dimen[1]{%
 \fontdimen8
 \ifx#1\displaystyle\textfont\else
 \ifx#1\textstyle\textfont\else
 \ifx#1\scriptstyle\scriptfont\else
 \scriptscriptfont\fi\fi\fi 3
}

\makeatother

\usepackage{graphicx}
\usepackage{placeins}

\usepackage{verbatim}

\usepackage{amssymb}
\usepackage[normalem]{ulem}
\usepackage{color}

\usepackage{hyperref}

\usepackage{hhline}
\usepackage{multirow}

\usepackage{float}
\usepackage{tabularx} 
\usepackage{array}

\usepackage[export]{adjustbox}
\usepackage{adjustbox}

\usepackage{ulem}
\newdimen\uulinesep
\usepackage[toc,page]{appendix}

\usepackage{makecell}

\newcommand{\mi}[4]{\ensuremath{\mat{{\mathcal{#1}}}{}^{(#2)#3}_{#4}}}

\newcommand{\ro}[3]{\ensuremath{\mat{\mathcal{#1}}{}^{#2}_{#3}}}
\newcommand{\roti}[3]{\ensuremath{\mat{\mathcal{\widetilde{#1}}}{}^{#2}_{#3}}}

\newcommand{\ti}[2]{\ensuremath{\mat{\mathcal{T}}{}^{#1}_{#2}}}

\newcommand{\lra}[1]{\ensuremath{\left(#1\right)}}

\newcommand{\ph}[1]{\phantom{#1}}

\newcommand{\ta}[3]{\setstacktabbedgap{#1}\parenMatrixstack[#2]{#3}}

\def\disp{\ensuremath\displaystyle}

\newcolumntype{x}[1]{>{\centering\arraybackslash\hspace{0pt}}p{#1}}

\usepackage{bbold}

\makeatletter
\newcommand{\specialnumber}[1]{%
 \def\tagform@##1{\maketag@@@{(\ignorespaces##1\unskip\@@italiccorr#1)}}%
}
\newcommand{\specialeqref}[2]{\begingroup
 \def\tagform@##1{\maketag@@@{(\ignorespaces##1\unskip\@@italiccorr#2)}}%
 \eqref{#1}\endgroup}
\makeatother

\usepackage{tabstackengine}
\stackMath

\usepackage{booktabs}

\usepackage{tasks}
\usepackage{amssymb}

\makeatletter
\def\p@subsection{}
\makeatother

\begin{document}

\author{S. Grytsiuk}
\email{s.grytsiuk@fz-juelich.de}
\affiliation{Peter Gr\"unberg Institut and Institute for Advanced Simulation, Forschungszentrum J\"ulich and JARA, 52425 J\"ulich, Germany}

\author{S. Bl\"ugel}
\affiliation{Peter Gr\"unberg Institut and Institute for Advanced Simulation, Forschungszentrum J\"ulich and JARA, 52425 J\"ulich, Germany}

\title{Micromagnetic description of twisted spin spirals in the B20 chiral magnet FeGe from first principles}

\begin{abstract}
Using the model of classical Heisenberg exchange and Dzyaloshinskii-Moriya (DM) interaction, we show that the ground state of the B20 FeGe chiral magnet is a superposition of twisted helical spin-density waves formed by different sublattices of the crystal. Such twisted spin-density waves propagate in the same direction but with different phases and different directions of the rotation axes. We derive an advanced micromagnetic expression describing the exchange and DM interaction for such magnetic structures. 
In particular, we show that such magnetic order gives rise to new contributions to the micromagnetic energies of the exchange and DM interactions.
By employing first-principles calculations based on density functional theory and using our micromagnetic model we show that the magnitude of the spin-spiral twist in B20 FeGe is of the same order as global spiraling. While the energy difference between the ground state of twisted spirals and the ferromagnetic state is in good agreement with the experimental results, for the spin spirals without a twist it is smaller by a factor of 3. In addition, we verify our results by employing spin-dynamics simulations. This calls for new experiments exploring the ground state properties of B20 chiral magnets.
\end{abstract}

\maketitle

\section{Introduction}

Non-centrosymmetric B20 chiral magnets have been reported to have the interesting property of breaking inversion symmetry, leading to various types of magnetic structures, ranging from spin spirals over skyrmions~\cite{Muhlbauer2009,Neubauer2009, Yu2010, Yu2011} and bobbers~\cite{Zheng2018,Rybakov2015} to the three-dimensional (3D) lattice of 3D magnetic textures~\cite{Tanigaki2015}, and thus holding great potential for innovative spintronic applications~\cite{Fert2013}. 
Due to the advent of different magnetic phases that can be tuned by temperature, magnetic fields, and other material parameters~\cite{Wilhelm2011, Pedrazzini2007, Shibata2015, Lee2009, Ritz2013, Deutsch2014, Rosler2012, Shibata2013, Grigoriev2013, Spencer2018, Huang2012}, the B20 systems not only host a rich phase diagram but their complex magnetic order and nontrivial topology in momentum space imprint also on unique transport and optical phenomena~\cite{Jonietz2011, Neubauer2009, Kanazawa2011, Chang2017, Takane2019, Yao2020, Sanchez2019}. 

Irrespective of the crucial relevance for explaining chiral magnetism in B20 compounds, however, a complete understanding of the underlying magnetic structures and magnetic interactions stabilizing them has been remarkably elusive.
While, for instance, the magnetic order at the ground state of many B20 compounds is often interpreted as a homogeneous helix~\cite{Ludgren1970, Bak1980} caused by relativistic Dzyaloshinskii-Moriya (DM) interactions~\cite{Dzyaloshinskii}, such a model ignoring higher-order magnetic interactions~\cite{Okumura2020} and temperature effects~\cite{Mendive2021} fails to explain the 3D texture of a few nanometers observed in B20 MnGe. Furthermore, a now well-known and well-accepted assumption made by Bak and Jensen 40 years ago~\cite{Bak1980} that DM interaction stabilizes a homogeneous helix in B20 materials is incomplete.
In particular, as shown by Chizhikov and Dmitrienko~\cite{Dmitrienko2012, Chizhikov2012, Chizhikov2013}, the DM vector that is perpendicular to the component responsible for helical spiraling gives rise to intersublattice canting in B20 compounds.
Such an effect can be viewed as the superposition of several helical spin-density waves in each sublattice propagating in the same direction but having different phases and different directions of the rotation axes (see Fig.~\ref{model}).

In this work, we derive an advanced micromagnetic energy equation describing the exchange and DM interactions for the spin spirals with twists formed by the magnetic moments of different sublattices in a crystal. 
While the micromagnetic energies of the exchange ($E^\text{ex}$) and DM interaction ($E^\text{dm}$) for the trivial (without twist) spin spirals have well-known dependencies on their wave vector  ${\bf q}$,
$$
\begin{array}{lcl}
     E^\text{ex}&=& E^\text{ex}(
     q^{0},
     {\bf q}^{\bigotimes{2}},\cdots),  \\
     E^\text{dm}&=& E^\text{dm}(
     {\bf q}^{1},
     {\bf q}^{\bigotimes{3}},\cdots),
\end{array}
$$
where ${\bf q}^{\bigotimes{p}}$ is the $p$-fold tensor product of ${\bf q}$ vector with itself, we show that the coupling between different sublattices gives rise to the twist of spin spirals with new contributions in ${\bf q}$ to the energies of magnetic interactions:
$$
\begin{array}{lcl}
     E^\text{ex}&=& E^\text{ex}(q^0,{\bf q}^1,{\bf q}^{\bigotimes{2}},{\bf q}^{\bigotimes{3}},\cdots),  \\
     E^\text{dm}&=& E^\text{dm}(q^0,{\bf q}^1,{\bf q}^{\bigotimes{2}},{\bf q}^{\bigotimes{3}},\cdots).
\end{array}
$$
Interestingly, the last expression implies also that even when $|{\bf q}|=q=0$, the ground state magnetic order might not be ferromagnetic (FM).
To substantiate the importance of such an effect in real materials, we employ density functional theory calculations for B20 FeGe since the energies without and with spin-orbit coupling in this compound are well described by the exchange and DM interactions~\cite{Grytsiuk2019}, in contrast to MnGe, for which higher-order interactions are expected to be pivotal~\cite{Grytsiuk2020}. We compute the atomistic interaction parameters via multiple-scattering theory as implemented in the Korringa-Kohn-Rostoker (KKR) Green's function method~\cite{kkr,Bauer2013, Papanikolaou2002}, and from them we determine the corresponding intersublattice micromagnetic parameters entering the advanced micromagnetic energy equation. By minimization of the micromagnetic energy, we determine the strength of global spiraling, as well as local twist between helices in different sublattices. 
We show that the magnitude of the spin spiral twist in B20 FeGe is of the same order as global spiraling and it reduces, by three times, the energy difference between the helical ground state and the FM state, resulting in good agreement with the experimentally measured saturation magnetic field.
Finally, we verify the results of our micromagnetic model by employing spin-dynamics simulations, that in addition to the twist of the spin spiral in B20 structures,  indicate their small nonhomogeneity.
However, since the energy gain due to the nonhomogeneity of the spin waves is much smaller than that due to their twist, for the sake of simplicity we ignore this effect in our micromagnetic model.

\section{Spin-spiral twist: Micromagnetic model}
\label{theory}

\begin{figure}[t!]\center
 \includegraphics[width=0.43\textwidth]{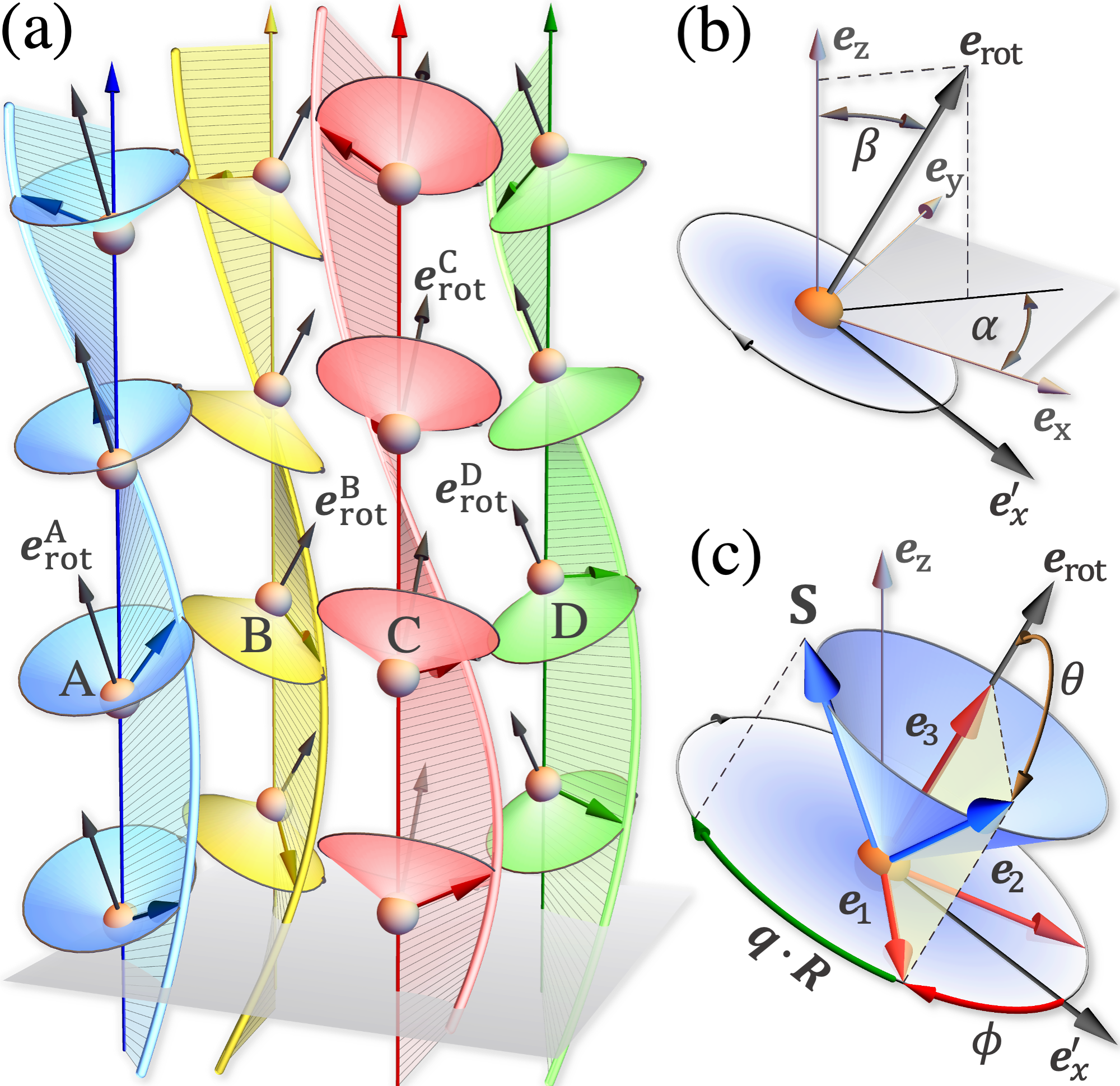}
 \caption{
 (a) Superposition of twisted spin-density waves formed by different sublattices ($A,B,C, \ldots $) of the crystal. Such spirals propagate in the same directions but with different (b) directions of the rotation axes  ${\bf e}_\text{rot}(\beta,\alpha)$ and (c) cone angles $\theta$ and phases $\phi$.
 The magnetic moment ${\bf S}$ in (c) is defined by cone angle $\theta$ with respect to 
 ${\bf e}_\text{rot} = \ro{R}{}{}(\beta,\alpha){\bf e}_z$ 
 and by phase $\phi$ and ${\bf q}\cdot {\bf R}$
 with respect to 
 ${\bf e}_x' = \ro{R}{}{}(\beta,\alpha){\bf e}_x$, where
 $\ro{R}{}{}(\beta,\alpha)$ is a rotation matrix mapping the ${\bf e}_z$ axis to the rotation axis ${\bf e}_\text{rot}$.
 A spin spiral with twist  $\Omega= \{\beta,\alpha,\theta,\phi\}$
 can be characterized by orthogonal vectors ${\bf e}_m(\Omega)$, $m=1,2,3$ [see Eq.~\eqref{ss}].
 }
 \label{model}
\end{figure}

We consider a one-dimensional spin-wave formed by conical homogeneous spin spirals in each sublattice $A$ of a crystal. For each spin spiral we assume the same wave vector ${\bf q}$ but a different rotation axis 
${\bf e}_\text{rot}^A(\beta^A,\alpha^A)$, cone angle $\theta^A$, and phase $\phi^A$ (see Fig.~\ref{model} and Appendix~\ref{sec:spirit}).
We define the orientation of any classical spin ${\bf S}_{i}^A$ of atom $i$ in sublattice $A$  as 
\begin{equation}
\label{ss}
 {\bf S}^A_i = 
 {\bf e}_1^A \cos ({\bf q}\cdot {\bf R}^A_i)
 + {\bf e}_2^A \sin ({\bf q}\cdot {\bf R}^A_i)
 + {\bf e}_3^A\,,
\end{equation}
where orthogonal vectors ${\bf e}_n^A = {\bf e}_n(\beta^A,\alpha^A,\theta^A, \phi^A)$ ($n=1,2,3$) characterize the spin-spiral twist in sublattice $A$ with a condition that $\vert{\bf S}_i^A\vert = 1$ (see Fig.~\ref{model}(c) and Appendix~\ref{rotation}). 
The upper index $A$ enumerates also the basis atom ($i=1$) with its atomic position ${\bf r}^A = {\bf R}_1^A$ in the chemical unit cell.  ${\bf R}_{i}^A = {\bf r}^A + \boldsymbol{\tau}_i$ defines the position of atom $i$ in sublattice $A$, where $\boldsymbol{\tau}_i$ is a translation vector.

We examine the effect of spin-spiral twist on the exchange and DM interactions following the effective spin-lattice Hamiltonian:
\begin{equation}
\label{en1}
E^{AB} =
-\dfrac{1}{N_A}
\displaystyle
\sum_{i=1}^{N_A}
\sum_{j=1}^{N_B} {\bf S}^A_i \,
\ro{J}{AB}{ij} \,
{\bf S}^B_j ,
\end{equation}
where the sum over $i$ is restricted to the number of magnetic atoms $N_A$ in sublattice $A$ of the magnetic unit cell and $j$ runs over atoms in sublattice $B$ of whole crystal structure. 
Tensor $\ro{J}{AB}{ij}$ represents the strength of the exchange ${J}_{ij}^{AB}$ and DM interaction  ${\bf D}_{ij}^{AB}$ between two sites $i$ and $j$ of sublattices $A$ and $B$, respectively:
\begin{equation}\label{jij}
\ro{J}{AB}{ij}= 
 \begin{pmatrix}
 \ph{-}{J} & \ph{-}D_{z} & -D_{y}\\
 -D_{z} & \ph{-}J & \ph{-}D_{x} \\
 \ph{-}D_{y} & - D_{x} & \ph{-}J 
 \end{pmatrix}_{ij}^{AB}.
\end{equation}

If the magnetic structure in each sublattice has characteristic length scales much larger than the underlying crystal lattice, then  the sum over $N_A$ (over the magnetic unit cell) in Eq.~\eqref{en1} can be approximated by the integral
\footnote{
Since
${\bf q}\cdot {\bf R}_i^A = 
{\bf q}\cdot({\bf r}^A + \boldsymbol{\tau}_i) = {\bf q}\cdot{\bf r}^A + {\omega_i}$
and
${\bf q}\cdot {\bf R}_j^B = 
{\bf q}\cdot({\bf R}^B_{j'} + \boldsymbol{\tau}_i) = {\bf q}\cdot{\bf R}^B_{j'} + {\omega_i}$, where ${\bf R}^B_{j'} = {\bf R}_j^B + \boldsymbol{\tau}_i$, $\boldsymbol{\tau}_i$ is translation vector (along the magnetic unit cell) and 
$\omega_i$ is an angle between two spins separated by $\boldsymbol{\tau}_i$,
the summation over the magnetic unit cell can be approximated by an integral
$\approx1/(2\pi) \int_0^{2\pi} \partial \omega{\bf S}({\bf q}\cdot {\bf r}^A +{\omega})\ro{J}{AB}{1j}{\bf S}({\bf q}\cdot {\bf R}^B_j +{\omega})$ assuming that $\omega_i \ll 2\pi$.
},
and the energy density of the magnetic interactions (between atoms in the chemical unit cell ${\bf r}^A$ and all other atoms ${\bf R}^B_j$) has a form
\begin{equation}
\label{en2}
\begin{array}{lr}
&E^{AB}({\bf q}) 
\approx \disp
-\frac{1}{2\pi} \sum_j^{N_B} \int_0^{2\pi}\hspace{-5px} \partial \omega \,
{\bf S}^A({\bf q} \cdot {\bf r}^{A} +\omega) \\
&\times \ro{J}{AB}{1j} 
{\bf S}^B({\bf q} \cdot {\bf R}_{j}^{B} + \omega) 
\end{array}
\end{equation}
Considering the magnetic structure given by Eq.~\eqref{ss}, the total energy after taking an integral in Eq.~\eqref{en2} has a form
\begin{equation}
\label{en3}
\arraycolsep=1pt\def\arraystretch{0}
\begin{array}{ll}
E^{AB}({\bf q}) \approx
\disp
-\sum_{j=1}^{N_B} 
\ro{J}{AB}{1j}\!:\!&\!
\bigg[\bigg.
\ro{C}{AB}{} +\disp \frac{1}{2}
\ro{C}{AB}{+}
\cos ({\bf q}\!\cdot\!{\bf R}_{1j}^{AB}) \\
&
+\disp \frac{1}{2}
\ro{C}{AB}{-}
\sin({\bf q}\!\cdot\!{\bf R}_{1j}^{AB})
\bigg.\bigg]
\end{array}
\end{equation}
where ${\bf R}_{1j}^{AB} = {\bf R}_{j}^{B} - {\bf r}^{A}$, the notation ``$:$" stands for the inner product of tensors of the same rank,  and tensors 
$\ro{C}{AB}{+} = {\bf e}_1^A \otimes {\bf e}_1^B + {\bf e}_2^A \otimes {\bf e}_2^B$, 
$\ro{C}{AB}{-} = {\bf e}_1^A \otimes {\bf e}_2^B - {\bf e}_2^A \otimes {\bf e}_1^B$, and 
$\ro{C}{AB}{} = {\bf e}_3^A \otimes {\bf e}_3^B$ characterize a twist between spin spirals in sublattices $A$ and $B$. By expanding cosine and sine
functions in Tailor series with respect to ${\bf q}\cdot {\bf R}$ (see Appendix~\ref{trig}) we arrive at the following micromagnetic expression for the total energy of spin spirals with twist:
\begin{equation}\label{en4}
\arraycolsep=1pt\def\arraystretch{1}
\begin{array}{lcl}
E^{AB}({\bf q}) &\approx&
 \mi{M}{0}{AB}{} :
 \ro{C}{AB}{} \\
&&+\disp \dfrac{1}{2}\sum_{p=0}\! 
\mi{M}{p}{AB}{}\!:\! 
\bigg(\ro{C}{AB}{p} \otimes \mi{Q}{p}{}{}\bigg),
\end{array}
\end{equation}
where the sum over $p$ defines the order of Tailor expansion, $\ro{C}{AB}{2p}=\ro{C}{AB}{+}$, $\ro{C}{AB}{2p+1}=\ro{C}{AB}{-}$, and $\mi{Q}{p}{}{} = {\bf q}^{\bigotimes{p}}$ is the $p$-fold tensor product of ${\bf q}$ vector with itself.
The tensor $\mi{M}{p}{AB}{}$ of rank $(p+1)$ characterizes the exchange and DM interactions between all sites of sublattices $A$ and $B$:
\begin{equation}
\label{micro1}
\mi{M}{p}{AB}{} = 
 \begin{pmatrix}
 \ph{-}\mi{A}{p}{}{} & 
 \ph{-}\mi{D}{p}{}{z} & 
 -\mi{D}{p}{}{y} \\
 -\mi{D}{p}{}{z} &
 \ph{-}\mi{A}{p}{}{} &
 \ph{-}\mi{D}{p}{}{x}\\
 \ph{-}\mi{D}{p}{}{y} &
 -\mi{D}{p}{}{x} &
 \ph{-}\mi{A}{p}{}{}
 \end{pmatrix}^{AB}.
\end{equation} 
The components of $\mi{M}{p}{AB}{}$ are tensors of rank $(p)$ and they are related to $\ro{J}{AB}{ij}$ and ${\bf R}^{AB}_{ij}$:
\begin{equation}
\label{kf1}
\arraycolsep=1pt\def\arraystretch{1}
\begin{array}{lcl}
%
\mi{M}{2p}{AB}{kk'} &=&
\disp 
\frac{(-1)^{p+1}}{(2p)!} 
\sum_{j=1}^{N_B} 
\ro{J}{AB}{1j(kk')}
\mi{R}{2p}{AB}{1j}\, ,
\\
%
\mi{M}{2p+1}{AB}{kk'}
&=&
\disp 
\frac{(-1)^{p+1}}{(2p+1)!} 
\sum_{j=1}^{N_B}
\ro{J}{AB}{1j(kk')}
\mi{R}{2p+1}{AB}{1j}\, ,
\end{array}
\end{equation}
where  $\mi{R}{p}{AB}{1j} = ({\bf R}^{AB}_{1j})^{\bigotimes{p}}$ is the $p$-fold tensor product of $ {\bf R}_{1j}^{AB}$ with itself (see Appendix~\ref{trig}). 
The micromagnetic tensors $\mi{A}{p}{AB}{}$ and $\mi{D}{p}{AB}{}=(\mi{D}{p}{AB}{x},\mi{D}{p}{AB}{y},\mi{D}{p}{AB}{z})$ 
in Eq.~\eqref{micro1} describe the micromagnetic exchange and DM interactions, respectively, between all ions of sublattices $A$ and $B$. 
Also, those tensors for different pairs of sublattices are related by symmetries of the underlying crystal structure and for B20 FeGe we list them in Appendix~\ref{sym}.

The micromagnetic energy given by Eq.~\eqref{en4} is a central equation of this work and it can be decomposed into the exchange and DM interactions (see Appendix~\ref{energies}).
From this equation it is clear that the twist of the spin spirals gives rise to the energies of the exchange and DM interactions, which are both functions of $\mi{Q}{p}{}{} = {\bf q}^{\bigotimes{p}}$, where $p=0,1,2,\cdots$.
This is in sharp contrast to trivial spin spirals, for which the exchange and DM interactions are functions  of only $\mi{Q}{2p}{}{}$ and $\mi{Q}{2p+1}{}{}$, respectively.

The main challenge is to find the twist tensors $\ro{C}{AB}{\pm}$ and $\ro{C}{AB}{}$ minimizing the total energy.
Using this model, we examine the twist of the spin-density waves in B20 FeGe, noting that it becomes difficult to carry out with full rigor, because of the 16 independent parameters ($\phi^A,\beta^A,\alpha^A, \theta^A$) characterizing the spin spiral’s twist in four sublattices ($A=1,2,3,4$) of the crystal. On the other hand, the proposed micromagnetic model has many attractive qualitative features since it gives solutions in terms of a few micromagnetic multisublattice parameters taking into account the symmetries of the crystal structure, whereas the atomistic model becomes too difficult to carry out  for a large supercell in which the magnetic structure is commensurate and due to the long-range character of the magnetic interactions in the itinerant magnets.
For our micromagnetic model, we assume homogeneous spin density waves in each sublattice of the crystal; as we demonstrate by means of spin-dynamics simulations (see Appendix~\ref{sec:spirit}), the gain in total energy due to the inhomogeneity of the spin-density waves is much smaller than that due to the twist.

\section{Computational details}

We compute the microscopic parameters of Heisenberg exchange $J_{ij}$ and DM ${\bf D}_{ij}$ interaction for B20 FeGe  by employing first-principles calculations via multiple-scattering theory as implemented in the KKR Green's function method~\cite{kkr,Bauer2013, Papanikolaou2002}.
In this framework, $J_{ij}$ and ${\bf D}_{ij}$ are obtained from the collinear state by applying infinitesimal rotations of the magnetic moments~\cite{Liechtenstein1987, Ebert2009}. 
All electronic structure calculations are performed for the experimental lattice parameter  and atomic positions of FeGe~\cite{Wappling1968, Lebech1989} in the local density approximation~\cite{Perdew92},
using $72\times72\times72$ ${\bf k}$-points in the full Brillouin zone with Fermi broadening of $100$K.
The micromagnetic parameters are computed using Eq.~\eqref{kf1},
for which the summation is truncated above a maximal interaction radius of $R_\text{max}=12a$, where $a$ is the lattice parameter.

\begin{figure}[t!]\center
 \includegraphics[width=0.475\textwidth]{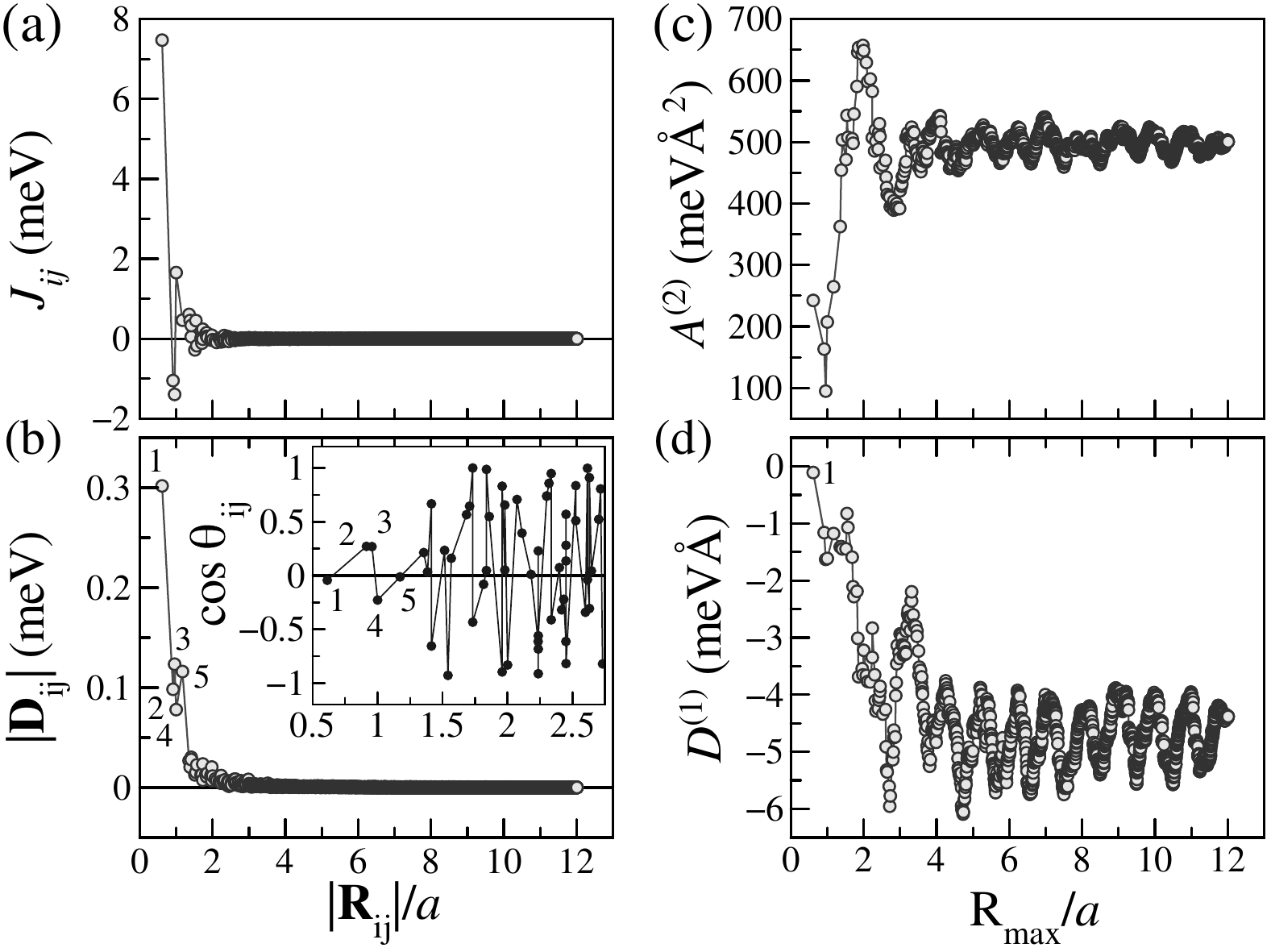}
 \vspace*{-3mm}
 \caption{
 (a) Exchange interaction parameters $J_{ij}$ and 
 (b) absolute values of DM interaction vectors 
 $\vert{\bf D}_{ij}\vert$ between Fe atoms 
 as a function of the interatomic distance (between Fe atoms) $\vert{\bf R}_{ij}\vert$ (in units of the lattice parameter $a$).
 (c) Micromagnetic spin stiffness $A^{(2)}$ and 
 (d) micromagnetic spiralization $D^{(1)}$ 
 as a function of $R_\text{max}$ (in units of the lattice parameter $a$), up to which contributions from the atomistic parameters ($J_{ij}$ and ${\bf D}_{ij}$ corresponding to all $\vert{\bf R}_{ij}^{AB}\vert<R_\text{max}$) are included. Note that parameters $J_{ij}$ and $\vert{\bf D}_{ij}\vert$ in (a) and (b) are multiplied by $\vert{\bf S}_i\vert\vert{\bf S}_j\vert$.
 The inset in (b) presents $\cos \theta_{ij}$ for the first few shells, where $\theta_{ij}$ is the angle between
 ${\bf R}_{ij}$ and ${\bf D}_{ij}$.
 }
 \label{dft1}
\end{figure}

\section{First-principles results and discussions}

\begin{figure*}[t!]\center
 \includegraphics[width=0.95\textwidth]{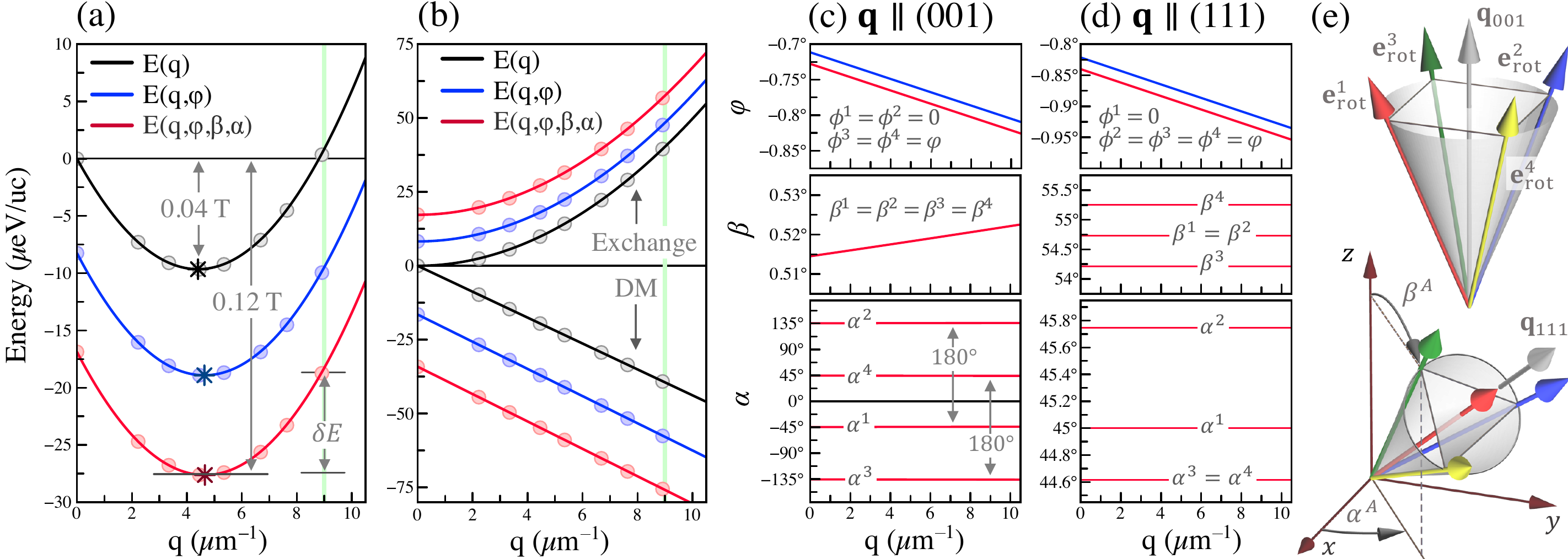}
 \vspace*{-2mm}
 \caption{
 (a) The total energy (per unit cell) of (b) the exchange and DM interactions, computed using the atomistic (circles) and micromagnetic (solid lines) models for the superposition of helical spin-density waves in B20 FeGe without twist (in black) and with twist (in blue and red) as a function of $q=|{\bf q}|$.
 The twist parameters of the spirals in each sublattice $A$ minimizing the total energy $\{\phi^A\}$ (blue curve) and $\{\phi^A,\beta^A,\alpha^A\}$ (red curves) for 
 (c) ${\bf q}_{001}$  and (d) ${\bf q}_{111}$.
 (e) Illustration showing the relative orientations of the rotation axes ${\bf e}_\text{rot}^A$ of the spin spirals in each sublattice $A$ for ${\bf q}_{001}$ and ${\bf q}_{111}$.
 Vertical arrows in (a) show the energy difference between the helical ground state and the FM state, corresponding to saturation magnetic fields of 0.04 and 0.12 T for spin spirals without and with twists, respectively. The solid green line stands for the experimental pitch $|{\bf q}^\text{exp}|=9\mu$m$^{-1}$.
 $\delta E$  is the total energy difference of the spin spiral with twist at ${\bf q}^\text{min}$ (the theoretical pitch) and ${\bf q}^\text{exp}$.
 }
 \label{dft2}
\end{figure*}

The atomistic parameters of the exchange interaction $J_{ij}$ and the absolute values of the DM interaction vectors $|{\bf D}_{ij}|$ as a function of the distance  $\vert {\bf R}_{ij}\vert$ between interacting pairs in B20 FeGe are shown in Figs.~\ref{dft1} (a) and (b).
While both atomistic interaction parameters decay fast with  $\vert {\bf R}_{ij}\vert$, their corresponding  micromagnetic parameters, $A^{(2)}= \frac{1}{2} \sum_{ij}J_{ij}R_{ij}^2$ and  $D^{(1)}=- \sum_{ij} {\bf R}_{ij}\cdot{\bf D}_{ij}$, respectively (for more details see Sec.~\ref{ss_identical}) have diminishing oscillatory behavior with respect to the summation cutoff  $|{\bf R}_{ij}|< R_\text{max}$, see Figs.~\ref{dft1} (c) and (d). 
The calculated value of the magnetic moment per Fe atom is ~$1.11~\mu_\text{B}$~\cite{Grytsiuk2019}, which is slightly larger than experimental value of $\sim 1.0$~$\mu_\text{B}$~\cite{Wappling1968, Ludgren1970, Spencer2018}.

\subsection{Spin wave formed by identical spirals} 
\label{ss_identical}

We first consider the simplest case in which the spin spirals in all sublattices are flat ($\theta^A=\pi/2$), have the same phase $\phi$ and the same direction of their rotation axes 
(${\bf e}^A_\text{rot} = {\bf e}_\text{rot} \parallel {\bf q}$).
In this case $\vert{\bf e}_3^A\vert = 0$, ${\bf e}_m^A \parallel {\bf e}_m^B\perp{\bf e}_\text{rot}$, and Eq.~\eqref{en4} after the summation over all pairs of sublattices in the B20 structure gives
\begin{equation}\label{en5}
\arraycolsep=1pt\def\arraystretch{1}
\begin{array}{lcl}
E_\text{I}({\bf q}) &=& \disp 
\sum_{p=0} \ro{A}{(2p)}{}{} \! :\! \mi{Q}{2p}{}{} +
\ro{D}{(2p+1)}{}{} \!:\! 
[\hat{\bf e}_\text{rot} \otimes \mi{Q}{2p+1}{}{} ]
\\
&=&\disp
A^{(0)} + D^{(1)} q + A^{(2)} q^2 + O_\text{I}({\bf q})
\end{array}
\end{equation}
where $q=\vert {\bf q}\vert$
and $O_\text{I}({\bf q})$ depicts the anisotropic higher-order contributions to the exchange and DM interaction, see Eq.~\eqref{en5ho}.
Tensors $\ro{A}{(2p)}{}{}=\sum_{AB}\ro{A}{(2p)AB}{}{}$ and $\ro{D}{(2p+1)}{}{}=\sum_{AB}\ro{D}{(2p+1)AB}{}{}$ characterize the micromagnetic exchange and DM interactions of order $p$, respectively, between magnetic ions in whole crystal.
Note that since $\ro{A}{(2p+1)AB}{}{}=-\ro{A}{(2p+1)BA}{}{}$ and $\ro{D}{(2p)AB}{}{}=-\ro{D}{(2p)BA}{}{}$, these odd contributions to the total energy after the summation over all pairs of sublattices for such trivial spin spirals cancel out. The energy contributions from the even terms $\ro{A}{(2p)AB}{}{}=\ro{A}{(2p)BA}{}{}$ and $\ro{D}{(2p+1)AB}{}{}=\ro{D}{(2p+1)BA}{}{}$ for such magnetic order remain.  For $p\leq2$ the obtained micromagnetic interactions tensors reduce to the identity matrices (see Appendix~\ref{sym}), 
\begin{align}
\arraycolsep=1pt\def\arraystretch{1}
\label{aa}
\ro{A}{(2)}{}{} & =  
 \dfrac{1}{2} \ro{I}{}{2}{} 
 \sum_{AB}
 \sum_{i\ne j} J_{ij}^{AB} |{\bf R}_{ij}^{AB}|^2 = A^{(2)}\ro{I}{}{2}{},\\
\label{dd} 
\ro{D}{(1)}{}{} &=  
-\displaystyle \ro{I}{}{2}{} 
\sum_{AB}
\sum_{i\ne j} {\bf D}_{ij}^{AB}\cdot{\bf R}_{ij}^{AB}= D^{(1)}\ro{I}{}{2}{}.
\end{align}

As follows from Eq.~\eqref{dd}, each bond ${\bf R}_{ij}^{AB}$ has the largest contribution to the spiralization parameter $D^{(1)}$ if ${\bf D}_{ij}^{AB} \parallel {\bf R}_{ij}^{AB}$, and it is zero if ${\bf D}_{ij}^{AB} \perp\ {\bf R}_{ij}^{AB}$.
Note that while $|{\bf D}_{ij}|$ are largest for the first few nearest neighbors, see Fig.~\ref{dft1}(b), their contribution to the total energy of the homogeneous flat spin spiral $\sim {\bf R}_{ij}\cdot{\bf D}_{ij} \sim \cos \theta_{ij}$, where $\theta_{ij}$ is an angle between ${\bf R}_{ij}$ and ${\bf D}_{ij}$, is negligibly small, see the inset in Fig.~\ref{dft1} (b) and Fig.~\ref{dft1} (d).
In contrast, as demonstrated by Chizhikov and Dmitrienko~\cite{Dmitrienko2012, Chizhikov2012, Chizhikov2013}, such normal components of the DM vectors, ${\bf D}_{ij} \perp\ {\bf R}_{ij}$, might give rise to the twist between spin spirals formed by different sublattices, which we discuss in the following sections.

While the lower-order contributions to the total energy,  Eq.~\eqref{en5}, of the exchange ($p=0,2$) and DM ($p=1$) interactions do not depend on the direction of the wave vector ${\bf q}$
\footnote{ Note, $\ro{A}{(2)}{}{}\!:\!\mi{Q}{2}{}{}\!=A^{(2)}\ro{I}{}{2}{}\!:\!\mi{Q}{2}{}{}=A^{(2)}q^2$ and for ${\bf e}_\text{rot}\parallel {\bf q}$  $\ro{D}{(1)}{}{}\!:\!\mi{Q}{1}{}{}\otimes {\bf e}_\text{rot}\!=D^{(1)}\ro{I}{}{2}{}:\mi{Q}{1}{}{}\otimes {\bf e}_\text{rot}\!=\!D^{(1)}q$, where $\ro{I}{}{p}{}$ is identity matrices of rank $p$.},
the higher-order contributions are anisotropic 
\begin{equation}\label{en5ho}
\arraycolsep=1pt\def\arraystretch{1}
\begin{array}{lcl}
O_\text{I}({\bf q}) &=& \disp 
D^{(3)}_1 
\begin{pmatrix}
q_x^2, q_y^2, q_z^2
\end{pmatrix}\cdot {\bf q}
+D^{(3)}_2 
\begin{pmatrix}
q_z^2,q_x^2,q_y^2
\end{pmatrix}\cdot {\bf q}
\\
&+&
D^{(3)}_3 
\begin{pmatrix}
q_y^2,q_z^2,q_x^2
\end{pmatrix}\cdot {\bf q} 
+
A^{(4)}_1 (q_x^4 + q_y^4 + q_z^4) \\
&+& A^{(4)}_2 (q_x^2 q_y^2 + q_x^2q_z^2 + q_y^2 q_z^2) + \cdots\, .
\end{array}
\end{equation}
The corresponding micromagnetic parameters for these higher-order terms, $D^{(3)}_n$ and $A^{(4)}_n$, are given in Appendix~\ref{sym}.
Since the contribution to the total energy of these terms for small values of $|{\bf q}|$ is small,  $|O_\text{I}({\bf q}_\text{min})| < 0.1$~($\mu$eV) $\ll |E_\text{I}({\bf q}_\text{min})| \approx 9$~($\mu$eV), we neglect them. Also, as follows from Fig.~\ref{dft2}(a),  energies obtained for ${\bf q}_{001}$ and ${\bf q}_{111}$ using Eq.~\eqref{en5} without $O_\text{I}({\bf q})$ contributions (black solid lines) are in good agreement with the atomistic model (black symbols) given by Eq.~\eqref{en1}.

\subsection{Spin spirals with a phase difference} 
\label{ss_phase}

Now we consider a case in which flat spirals ($\theta^A=\pi/2$) in each sublattice $A$ are allowed to have different phases $\phi^A$ while the directions of the rotation axes ${\bf e}_\text{rot}^A = {\bf e}_\text{rot}(\beta,\alpha) = {\bf q}/\vert {\bf q}\vert$ remain the same. 
For such a choice of the magnetic structure the total energy of the magnetic interactions between atoms in sublattices $A$ and $B$, Eq.~\eqref{en4}, has the form
\begin{equation}\label{phase1}
\arraycolsep=1pt\def\arraystretch{1}
\begin{array}{rlcr}
E^{AB}_\text{II}({\bf q}, \!&\!\phi^{AB}) = &
\cos \phi^{AB} \sum_p \Big( \mi{A}{2p}{AB}{}: \mi{Q}{2p}{}{} \Big.\\
&&+\mi{D}{2p+1}{AB}{} :\mi{Q}{2p+1}{}{} \otimes {\bf e}_\text{rot} \Big.\Big)\\
+&\sin \phi^{AB} & \sum_p
\Big(\!-\!\mi{A}{2p+1}{AB}{}: \mi{Q}{2p+1}{}{} \Big.\\
&&+\mi{D}{2p}{AB}{} 
: \mi{Q}{2p}{}{} \otimes {\bf e}_\text{rot} \Big.\Big)
\end{array}
\end{equation}
where $\phi^{AB} = \phi^{B} -\phi^{A}$ is the phase difference between spin spirals in sublattices $A$ and $B$.
Note that if $\phi^{AB}=0$, then  $E_{II}^{AB}({\bf q}) = E_{I}^{AB}({\bf q})$, see Eq.~\eqref{en5}. As follows from Eq.~\eqref{phase1}, the phase difference $\phi^{AB}$ corresponding to the lowest total energy of the magnetic interactions between two sublattices 
$$
\phi^{AB}_\text{min}({\bf q})  = \arctan \left[ \dfrac{\mi{D}{0}{AB}{} \cdot {\bf e}_\text{rot} - \mi{A}{1}{AB}{} \cdot {\bf q} + \cdots
}{\mi{A}{0}{AB}{} + {\bf e}_\text{rot}^T\cdot \mi{D}{1}{AB}{} \cdot {\bf q} + \cdots
}\right]
$$
is nonzero if 
$\mi{D}{2p}{AB}{}$ or $\mi{A}{2p+1}{AB}{}$ are nonzero tensors.

\begin{figure}[t!]\center
 \includegraphics[width=0.43\textwidth]{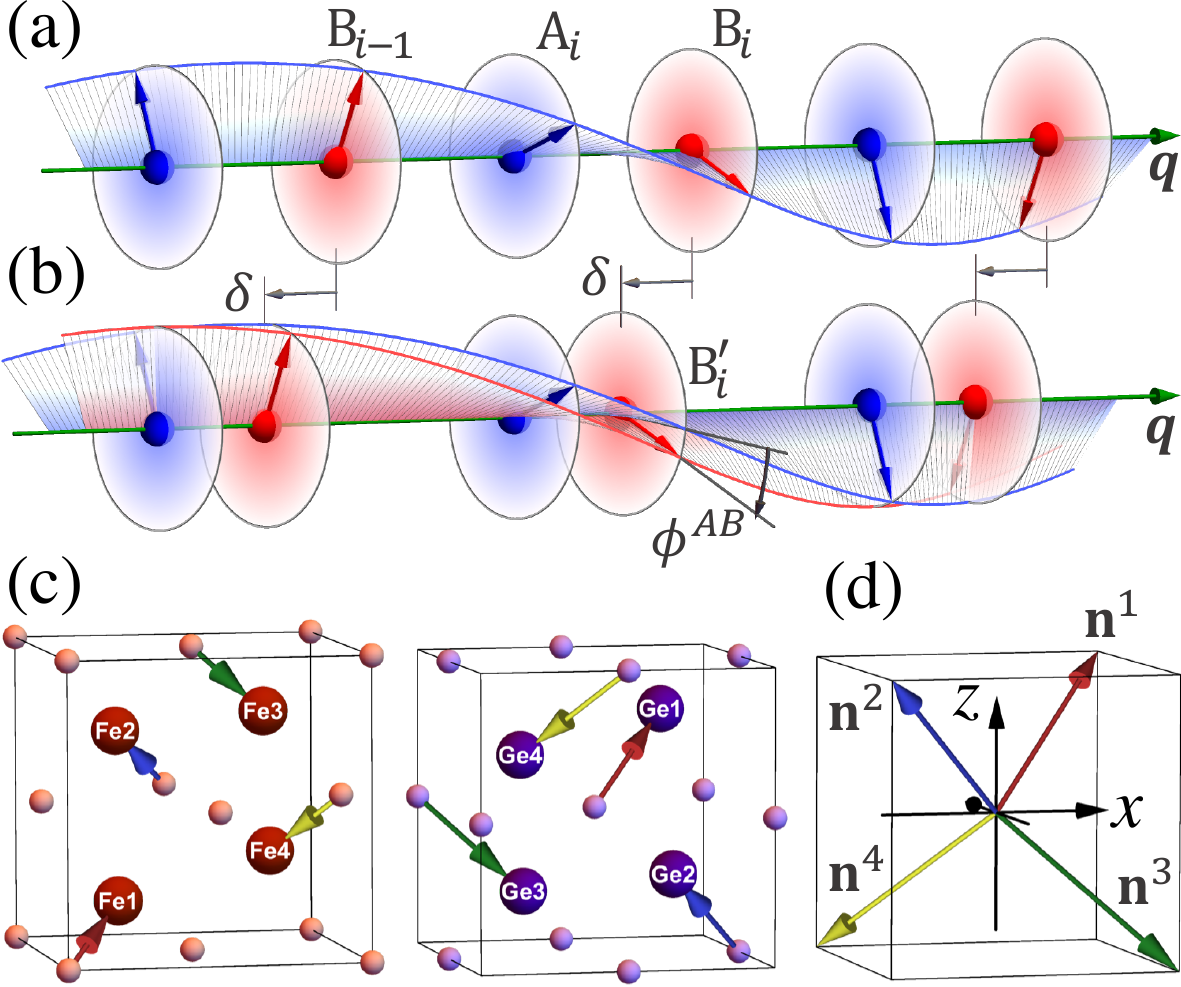}
 \caption{
 (a) and (b) Illustrations of the spin-density wave formed by spin spirals in sublattices $A$ and $B$. 
 Spin spirals in (a) are identical, and those in (b) have phase difference $\phi^{AB} = \boldsymbol{\delta} \cdot {\bf q}$ 
 due to the displacement of atoms in sublattice $B\rightarrow B'$ by $\delta$ such that ${\bf S}_i^B = {\bf S}_i^{B'}$.
 (c) Displacements of Fe and Ge ions from high-symmetry sites of the NaCl structure (smaller spheres) to lower-symmetry sites of the B20 structure (larger spheres) along cube diagonals ${\bf n}^A$, as shown in (d).
 }
\label{model2}
\end{figure}

To explain the emergence of the phase difference let us assume first only two spin spirals formed by sublattices $A$ and $B$.  When each atom $i$ in sublattices $A$ and $B$ possesses an inversion symmetry with respect to their relative positions, ${\bf R}_{ii}^{AB} = - {\bf R}_{ii-1}^{AB}$, and magnetic interactions, $\ro{J}{AB}{ij} = (\ro{J}{AB}{ii-1})^T$, as shown in Fig.~\ref{model2}(a), then the summation in Eq.~\eqref{kf1} gives $\mi{D}{2p}{AB}{} = \mi{A}{2p+1}{AB}{} = 0\ro{I}{}{2p+1}{}$; therefore, $\phi_\text{min}^{AB}({\bf q})=0$.
However, when atoms in one of the sublattices are displaced from their symmetric positions by $\boldsymbol{\delta}$, see Fig.~\ref{model2} (b), then $\phi_\text{min}^{AB} ({\bf q}) = \boldsymbol{\delta}\cdot {\bf q}$ if $\ro{J}{AB}{ii} = (\ro{J}{AB}{ii-1})^T$ and the magnetic moments remain the same directions as before the displacement (${\bf S}_i^B = {\bf S}_i^{B'}$). If $\ro{J}{AB}{ij}\neq (\ro{J}{AB}{ii-1})^T$, then ${\bf S}_i^B \neq {\bf S}_i^{B'}$, and $\phi_\text{min}^{AB} ({\bf q})\neq \boldsymbol{\delta}\cdot {\bf q}$ in general. 

Similarly, the emergence of the relative phases $\phi^{AB}$ between spin spirals in different sublattices of the B20 structure is expected because atoms in this structure are displaced with respect to the high-symmetry sites [given by space group $Fm3m$, see Fig.~\ref{model2}(c)].
Such displacement of atoms in each sublattice $A$ can be defined as $\boldsymbol{\delta}^A = {\bf n}^A u^{M}$, where the parameter $u^\text{M}$ stands for the Wyckoff positions of magnetic ions in the B20 structure (space group $P2_13$) and ${\bf n}^A$ is the symmetry direction at each site in sublattice $A$ (see Appendix~\ref{sym}).
Finding phase $\phi^{A}$ for each spiral $A$ in a B20 chiral magnet for a given direction of ${\bf q}$ requires taking into account the magnetic interactions between all pairs of sublattices.
Since only a phase difference $\phi^{AB}$ enters Eq.~\eqref{phase1}, we are allowed to choose $\phi^{1}=0$. 
Assuming ${\bf e}_\text{rot}\parallel {\bf q}$ and minimizing (numerically) the total energy, Eq.~\eqref{phase1}, for ${\bf q}_{001}$ and ${\bf q}_{111}$ with respect to the phases $\phi^{A}$, we obtain
\begin{equation}\label{phi_min}
 \begin{array}{ll}
 & \phi^1=\phi^2=0,\, \phi^3=\phi^4=\varphi_{001} ({\bf q}_{001}),\\
 & \phi^1=0,\, \phi^2=\phi^3=\phi^4=\varphi_{111} ({\bf q}_{111})\\
 \end{array}
\end{equation}
Therefore, due to the symmetries of the B20 structure the only unknown parameter $\varphi({\bf q})$ remains for the two cases of ${\bf q}$.
The above equations indicate also that the phase difference $\phi^{AB}$ between spirals in sublattices $A$ and $B$ can be defined in terms of the relative atomic displacements from the high-symmetry sites projected on 
${\bf e}_\text{rot}\parallel {\bf q}$:
$$
\phi_\text{min}^{AB} ({\bf q})= 
{\bf e}_\text{rot}
\cdot ({\bf n}^B-{\bf n}^A) \,
\varphi({\bf q})
$$
Since, due to the symmetries, the only unknown parameter $\varphi({\bf q})$ remains, the total energy in Eq.~\eqref{phase1} can be easily minimized, and  $\varphi_\text{min}({\bf q})$  that includes the interactions between magnetic atoms in all sublattices of the crystal can be computed.
The energies of the exchange and DM interactions as a function of $\vert{\bf q}\vert$ and $\varphi_\text{min}({\bf q})$ for ${\bf q}_{001}$ or ${\bf q}_{111}$ are shown as blue circles (atomistic model) and blue lines (micromagnetic model) in Figs.~\ref{dft2}(a) and (b).  The corresponding $\varphi_\text{min}({\bf q})$ for ${\bf q}_{001}$ and ${\bf q}_{111}$ are shown in Figs.~\ref{dft2}(c) and \ref{dft2}(d), respectively.
As follows from Fig.~\ref{dft2} the energies corresponding to $\varphi_\text{min}({\bf q})$ for the two directions of ${\bf q}$ are almost the same (a small difference is due to the higher-order contributions; see Sec.~\ref{ss_identical}). 
The obtained energies are lower than in the case of $\varphi({\bf q})=0$,
$
E^{AB}_\text{II}({\bf q}_\text{min}, \varphi_\text{min})<
E^{AB}_\text{I}({\bf q}_\text{min})
$,
as the energy gain of the DM interaction dominates the energy loss of the exchange contribution, see Fig~\ref{dft2} (b).
Also, the obtained phase difference $\varphi_\text{min}({\bf q}_\text{min})\approx -0.8^\circ$, is of the same order as the angle between magnetic moments of nearest atoms formed due to the global spiraling of $\approx 1.28^\circ$
\footnote{The angle between magnetic moments of nearest atoms formed due to the global spiraling can be defined as $\approx 2\pi a/(4 \lambda) = 1.28^\circ$, where $a$ is lattice parameter, $\lambda \approx 70 a$ is an experimental period of the spin-spiral modulation and factor $4$ counts 4 magnetic ions in the unit cell.}.

\subsection{Spirals with different cones} 
\label{ss_cone}

\begin{figure}[t!]\center
 \includegraphics[width=0.43\textwidth]{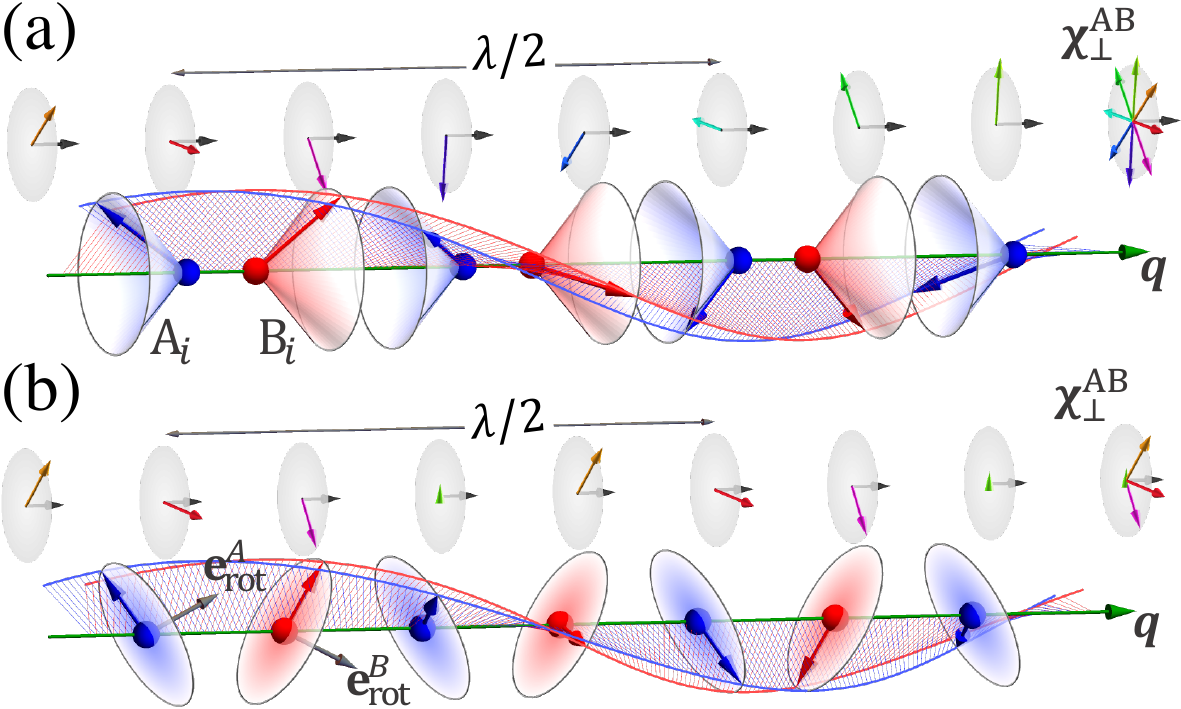}
 \caption{Illustration of the superposition of spin spirals $A$ and $B$ with:
 (a) different cone angles and (b) different orientations of the rotation axes.
 The rainbow colors for the vectors represent different orientations of the vector chirality 
 $\boldsymbol{\chi}^{AB}_{i} = {\bf S}^A_i\times {\bf S}^B_i$ component normal to ${\bf q}$.
 }
\label{model3}
\end{figure}

Now we consider a case in which in addition to the phase difference $\phi^{AB}$ between spin spirals each of them is allowed to have a cone angle $\theta^A$ between each ${\bf S}_i^A$ and  ${\bf e}_\text{rot}^A$ that might differ from $\pi/2$, see Fig.~\ref{model3}~(a). The total energy of the magnetic interactions for such magnetic structure has the form
\begin{equation*}
\arraycolsep=1pt\def\arraystretch{1}
\begin{array}{lcl}
E^{AB}_\text{III}({\bf q},\phi^{AB}, \theta^A,\theta^B) &=& 
\cos \theta^A \cos \theta^B E_\text{FM}^{AB} \\
& +&\sin \theta^A \sin \theta^B E^{AB}_\text{II}({\bf q}, \phi^{AB})
\end{array}
\end{equation*}
where $E_\text{FM}^{AB}= - \sum_{ij} J_{ij}^{AB} = \mi{A}{0}{AB}{}$ and $E^{AB}_\text{II}({\bf q}, \phi^{AB})$ are the total energies, respectively, of the FM state and spin-spiral state with phase difference $\phi^{AB}$, see Eq.~\eqref{phase1}.
In such a case the chirality vectors $\boldsymbol{\chi}_{ij}^{AB}={\bf S}_i^A\times {\bf S}_j^B$ of the spin-spiral state have components that are normal and parallel to the direction of the wave vector ${\bf q}$, $\boldsymbol{\chi}_{ij}^{AB} = \boldsymbol{\chi}_{ij\parallel {\bf q}}^{AB} + \boldsymbol{\chi}_{ij\perp {\bf q}}^{AB}$, see Fig.~\ref{model3}~(a).
Case $\theta^A=\theta^B=0^\circ$ ($\theta^A=\theta^B=90^\circ$) corresponds to the FM (flat spin spiral) state.
Any deviations of the cone angles $\theta^A$ and $\theta^B$ from $\pi/2$ reduces the energy contribution of $E^{AB}_\text{II}({\bf q}, \phi^{AB})$ and the total energy is in between the total energy of the FM state and the total energy of the flat spin spirals.
This is in particular because conical spin spirals reduce the contribution to the total energy of the DM vectors parallel to ${\bf q}$, as ${\bf D}_{ij}\cdot \boldsymbol{\chi}_{ij\parallel {\bf q}}^{AB}= \sin \theta^A \sin \theta^B {\bf D}_{ij}\cdot \boldsymbol{\chi}_{ij}^{AB}$.
Also, the components of DM vectors normal to ${\bf q}$ on average over the magnetic unit cell give zero contribution to the total energy, as vector chirality components  $\boldsymbol{\chi}_{ij\perp {\bf q}}^{AB}$ rotate by $2\pi$ over the whole magnetic unit cell, see Fig.~\ref{model3} (a). From this, we conclude that the competition between the exchange and DM interactions favors the emergence of flat spin spirals rather than those with $\theta^A \neq \pi/2$.

\subsection{Spirals with different rotation axes} 
\label{ss_rot}

Finally, we consider a case in which flat spirals in each sublattice in addition to the phase difference $\phi^{AB}$ are allowed to have different orientations of their rotation axes ${\bf e}^A_\text{rot}$, 
characterized by angles $\beta^A$ and $\alpha^A$ as  illustrated in Fig.~\ref{model}.
Like in the case discussed above, different orientations of the rotation axes in two sublattices give rise to $\boldsymbol{\chi}_{ij\parallel {\bf q}}^{AB}$ and $\boldsymbol{\chi}_{ij\perp {\bf q}}^{AB}$ vector chirality components, see Fig.~\ref{model3} (b).
While the component $\boldsymbol{\chi}_{ij\parallel {\bf q}}^{AB}$ caused by the DM interaction vectors parallel to ${\bf q}$ gives rise to the global spiraling, the components $\boldsymbol{\chi}_{ij\perp {\bf q}}^{AB}$ might arise due to  the DM interaction vectors ${\bf D}_{ij\perp {\bf q}}^{AB}$ normal to ${\bf q}$. 
Although the contribution of DM interaction from  $\boldsymbol{\chi}_{ij\perp {\bf q}}^{AB}$ varies over the magnetic unit cell,  it does not vanish as $\boldsymbol{\chi}_{ij\perp {\bf q}}^{AB}$ rotates around ${\bf q}$ by only $\pi$. 

The directions of the spin-spirals rotation axes formed by different sublattices of the B20 structure can be obtained from the minimization of the total energy, Eq.~\eqref{en4}, for a given wave vector ${\bf q}$.  In this case the total energy of the magnetic interactions is a function of $13$ parameters (${\bf q}$, $\phi^A$, $\beta^A$, and $\alpha^A$ for $A=1,2,3,4$).  As we will demonstrate, the number of unknown parameters can be reduced due to symmetries, but let us first briefly discuss the case $\vert{\bf q}\vert=0$, for which the total energy given by Eq.~\eqref{en4} has the form 
\begin{equation}
\arraycolsep=1pt\def\arraystretch{1}
\begin{array}{ll}
E^{(0)}_\text{IV} &=\disp
\frac{1}{2} \disp \sum_{AB}
\Big(
{\bf e}_1^A \cdot {\bf e}_1^B +
{\bf e}_2^A \cdot {\bf e}_2^B
\Big) 
\mi{A}{0}{AB}{} \\
&+\disp
\frac{1}{2} \sum_{AB} 
 \Big(
 {\bf e}_1^A \times {\bf e}_1^B +
 {\bf e}_2^A \times {\bf e}_2^B
 \Big) \cdot
\mi{D}{0}{AB}{} 
\end{array}
\end{equation}
From this equation it follows that the contribution from the DM interaction takes maximal values when ${\bf e}_m^A \perp {\bf e}_m^B \perp \mi{D}{0}{AB}{}$. Note that in the case with ${\bf e}_\text{rot}^A = {\bf e}_\text{rot}^B$ we have $({\bf e}_m^A \times {\bf e}_m^B) = {\bf e}_\text{rot}^A\sin \phi^{AB}$, and only $\mi{D}{0}{AB}{}\parallel {\bf e}_\text{rot}^A$ contributes to the total energy for $|{\bf q}|=0$ (see Sec.~\ref{ss_phase}). However, since $\mi{D}{0}{AB}{} \nparallel \mi{D}{0}{AC}{}$ in B20 FeGe (see Table~\ref{sym3} in Appendix~\ref{sym}), the DM interaction, in addition to the phase $\phi^{AB}$, might lead to ${\bf e}_\text{rot}^B \nparallel {\bf e}_\text{rot}^C$.

The energies (neglecting small higher-order anisotropic contributions) of the magnetic interactions in B20 FeGe minimized numerically with respect to the twist angles ($\phi^A$, $\beta^A$, and $\alpha^A$) as a function of ${\bf q}_{001}$ and ${\bf q}_{111}$ are shown as red lines in Fig.~\ref{dft2}(a) and \ref{dft2}(b). While the minimized total energy does not depend on the direction of ${\bf q}$, the corresponding twist's angles of the spirals in each sublattice $A$ depend on it, such that 
$$
\begin{array}{ll}
&\disp 
\sum_A^4 {\bf n}^A \cdot {\bf e}_\text{rot}^A = 0, \quad
\big(\sum_A^4 {\bf e}_\text{rot}^A \big) \parallel {\bf q}\\
&\disp 
\phi_\text{min}^{AB} ({\bf q}) =
({\bf n}^B\cdot {\bf e}_\text{rot}^B -
{\bf n}^A \cdot {\bf e}_\text{rot}^A) \,
\varphi({\bf q})
\end{array}
$$
As the result, only three parameters ($\phi,\beta,\alpha$) out of 12 $\{\phi^A,\beta^A,\alpha^A\}$ remain independent. As an example, in the case of ${\bf q}_{001}$ we obtain
\begin{equation}\label{erot_min}
\begin{array}{ll}
&
\phi^1=\phi^2 = 0,\, \quad \phi^3=\phi^4=\phi(q),\\
&
\beta^1=\beta^2=\beta^3=\beta^4 = \beta(q),\\
&
\begin{array}{lcrcr}
\alpha^{1} &=& \alpha(q),\quad 
\alpha^{2} &=& \alpha(q)+\pi,\\
\alpha^{4} &=& -\alpha(q),\quad 
\alpha^{3} &=& -\alpha(q)-\pi,
\end{array}
\end{array}
\end{equation}
A similar result can be shown for ${\bf q}_{111}$. In this case 
${\bf e}^1_\text{rot} \parallel {\bf q} \parallel {\bf n}^1$, 
${\bf e}^2_\text{rot} \cdot {\bf q} = {\bf e}^3_\text{rot} \cdot {\bf q} = {\bf e}^4_\text{rot} \cdot {\bf q}$ and ${\bf e}^2_\text{rot} \cdot {\bf n}^2 = 
{\bf e}^3_\text{rot} \cdot {\bf n}^3 = {\bf e}^4_\text{rot} \cdot {\bf n}^4$.
The orientations of the rotation axes for the spin spirals in each sublattice of B20 FeGe for ${\bf q}_{001}$ and ${\bf q}_{111}$ are shown in Fig.~\ref{dft2}(e).
The energy difference of about $0.028$~meV between the helical ground state with twist and the FM state is three times lower than in the case without twist, corresponding to a saturation magnetic field of $B=0.12$~T
~\footnote{The saturation magnetic field ${\bf B}\parallel {\bf q}$ corresponds to the Zeeman energy  $E_\text{Z}({\bf q}_\text{min}) = -{\bf m}\cdot {\bf B}= E({\bf q}_\text{min}) - E_\text{FM}$, where $|{\bf m}|= 4\mu_B = 4\times 57.8838\, (\mu eV T^{-1})$ is the total magnetic moment in the unit cell. Since the saturation magnetic filed leads to ${\bf m}\parallel {\bf B}$ we get $|{\bf B}| = -E_Z({\bf q}_\text{min})/|{\bf m}|$}, which is in good agreement with the experimental value obtained for FeGe~\cite{Bauer2016, Spencer2018, Ludgren1970}.

\section{Conclusions}

In this work, we studied the magnetic interactions between  sublattices of non-centrosymmetric crystal using the micromagnetic model. In particular, we showed that due to the exchange and DM interactions the ground state spin spirals formed by sublattices of B20 FeGe have different phases and different orientations of their rotation axis (see Eq.~\eqref{ss}).  We developed a micromagnetic model describing the exchange and DM interactions for such twisted spin-density waves (Eq.~\eqref{en4}), which gives a solution in terms of a few micromagnetic multi-sublattice interaction parameters taking into account symmetries of the crystal structure, whereas the atomistic model becomes too difficult to carry out due to the long range of the magnetic interactions.  To uncover the importance of the spin-spiral twist for the energetics of the materials with multi-sublattices without inversion symmetry we compute the micromagnetic exchange and DM interaction parameters for B20 FeGe, as other higher-order interactions in this compound are expected to be small~\cite{Grytsiuk2019, Grytsiuk2020}. From the minimization of the micromagnetic total energy, we obtain the ground state parameters characterizing the twist of the spin-density waves in each sublattice of the crystal. In particular, we show that the magnitude of this effect in B20 FeGe is of the same order as global spiraling and, it lowers the total energy (compare to the FM state)  three times compared to the case when it is ignored.

While the twist of the spin spirals in B20 FeGe significantly reduces the total energy of the exchange and DM interactions, the obtained period of the magnetic modulations $\lambda_\text{min}^\text{DFT} = 2\pi/q_\text{min}^\text{DFT}$ remains two times larger than the experimental observation $\lambda_\text{min}^\text{exp} = 2\pi/q^\text{exp}_\text{min}$, see Fig.~\ref{dft2}(a).
A possible reason for this discrepancy might stem from the failure of DFT to capture the effect of electronic correlations which can dramatically change the overall behavior of the system, as was recently demonstrated 
using dynamical mean-field theory~\cite{Borisov2020}.
While the exchange interaction parameters $J_{ij}$ obtained by DFT explain  the experimental Curie temperature of FeGe very well ~\cite{Grytsiuk2019},  the electronic correlations might have a stronger impact on the strength of the DM interaction.  
As shown in Fig.~\ref{fi:scale} in Appendix~\ref{sec:spirit}, not  only does the period of the spin spiral become shorter when the strength of the DM interaction is increased, but also the corresponding twist angles $\phi_\text{min}$ and $\beta_\text{min}$ become larger, lowering the total energy more prominently compared with the case when the twist is ignored. On the other hand, since the difference in total energy for ${\bf q}_\text{min}^\text{DFT}$ and ${\bf q}^\text{exp}_\text{min}$ due to the twist is small, $E({\bf q}_\text{min}^\text{DFT},\Omega)-E({\bf q}_\text{min}^\text{exp},\Omega) = 8\mu$eV (see Fig.~\ref{dft2}(a)), we speculate that it can be overtaken by higher-order magnetic interactions~\cite{Grytsiuk2020} or by quantum fluctuations~\cite{Povzner} shifting ${\bf q}_\text{min}^\text{DFT}$ towards ${\bf q}_\text{min}^\text{exp}$.

\begin{acknowledgments}
We thank F. Rybakov and N.~Kiselev for  fruitful discussions. 
We acknowledge financial support from the DARPA TEE program through a MIPR grant (\#HR0011831554) from DOI and from Deutsche Forschungsgemeinschaft (DFG) through SPP 2137 ``Skyrmionics" (Project BL 444/16) and the Collaborative Research Centers SFB 1238 (Project C01), as well as computing resources at the supercomputers JURECA at J\"ulich Supercomputing Centre and JARA-HPC from RWTH Aachen University (Projects jara0224 and jara3dmagic). 
\end{acknowledgments}

\appendix

\section{Spin-dynamic simulations}
\label{sec:spirit}

\begin{figure}[t!]\center
 \includegraphics[width=0.48\textwidth]{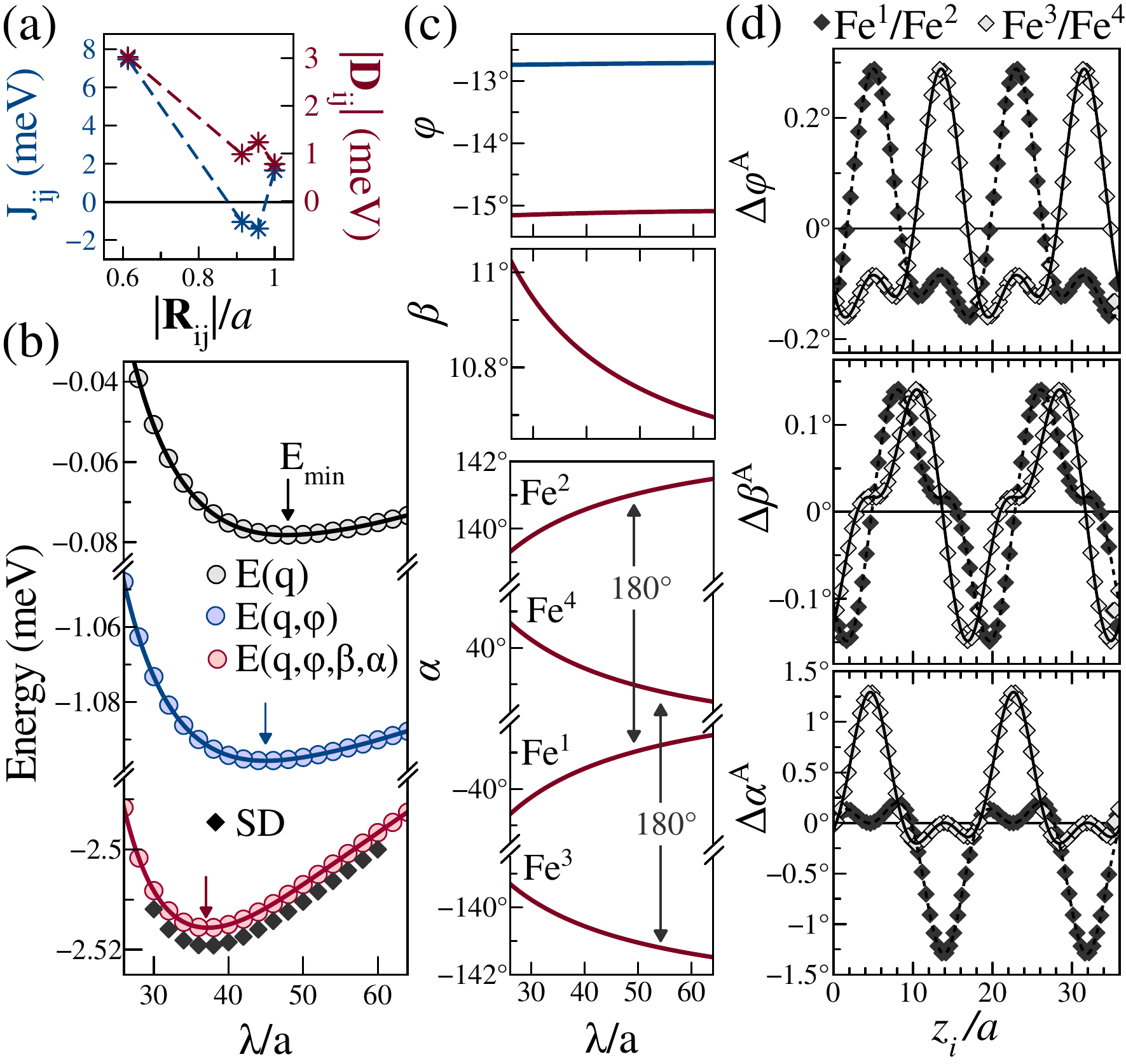}
 \vspace*{-2mm}
 \caption{
 (a) Exchange interaction parameters $J_{ij}$ and absolute values of DM interaction vectors $|{\bf D}_{ij}|$ between Fe atoms as a function of the interatomic distance $|{\bf R}_{ij}|$ (in units of the lattice parameter $a$). (b) The total energy of the exchange and DM interaction, computed using the micromagnetic model (solid lines), atomistic model (circles), and spin-dynamics (SD) simulations (diamonds) for the spin spiral propagating along the $(001)$ direction as a function of 
 $\lambda = 2\pi/\vert{\bf q}\vert$.
 The black curve stands for the spin spirals without twist.
 Blue and red curves represent twisted spin spirals characterized by
 $(\varphi)$  and $(\varphi, \alpha^A, \beta)$, respectively, 
 see Eqs.~\eqref{phi_min} and \eqref{erot_min}.
 (c) Twist parameters $\varphi$, $\beta$, and $\alpha^A$ minimizing the micromagnetic total energy as a function of $\lambda$. 
 (d) Difference between the twist parameters obtained from the spin-dynamics simulations 
 $\Omega^A_i=\{\phi_i^A,\beta_i^A, \alpha_i^A\} = \Omega^A_i(z_i^A)$ and twist parameters obtained from the micromagnetic model $\Omega^A=\{\phi^A,\beta^A, \alpha^A\} = \text{const}$ as a function of the atomic position along the $z$-direction in a $1\times1\times 36$ magnetic unit cell.
 }
 \label{fi:spirit}
\end{figure}

\begin{figure}[t!]\center
 \includegraphics[width=0.48\textwidth]{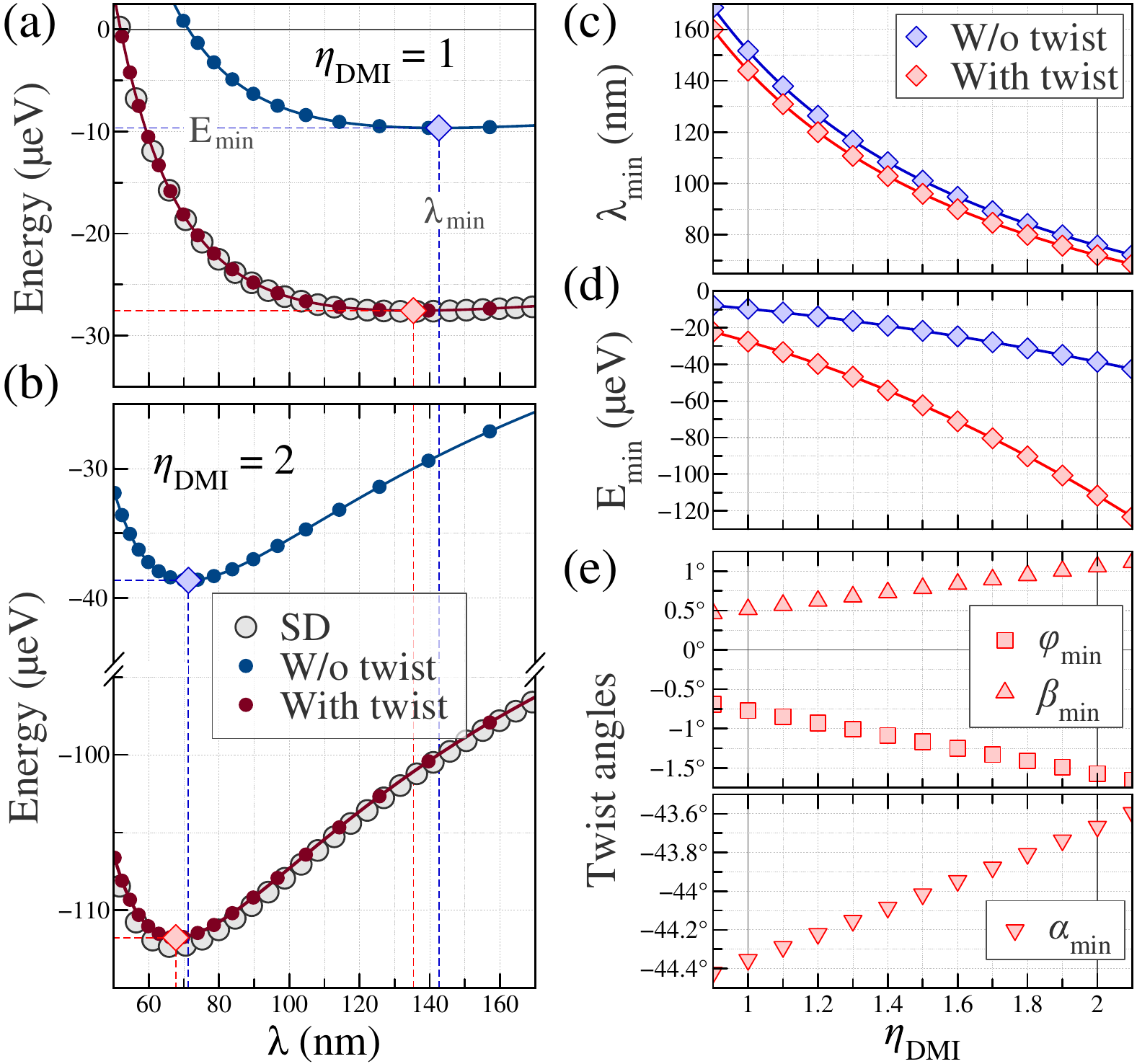}\vspace*{-2mm}
 \caption{
(a) and (b) The total energy of the exchange and DM interaction as a function of spin spiral period $\lambda$, computed for ${\bf q} \parallel (001)$ using scaling factors $\eta_\text{DMI}=1$ and $\eta_\text{DMI}=2$, respectively, for the length of the DM interaction vectors  ${\bf D}_{ij}$ of FeGe. The energies of the SD simulations (large gray circles) and atomistic model (small circles) are obtained using the effective $\roti{J}{AB}{ij}$ and real $\ro{J}{AB}{ij}$ interaction parameters, respectively. The micromagnetic energies are minimized for spin spirals without twist (blue lines) and with a twist (red lines). The diamonds  indicate $\lambda_\text{min}$ corresponding to the minimum of the total energy $E_\text{min}$.
(c) Period $\lambda_\text{min}$,  (d) energy minimum $E_\text{min}$, and (e) corresponding twist angles ($\phi_\text{min}$, $\beta_\text{min}$, $\alpha_\text{min}$) of the spin spirals with twist as a function of the scaling factor $\eta_\text{DMI}$ for the length of ${\bf D}_{ij}$.
 }
 \label{fi:scale}
\end{figure}

Here, by employing atomistic spin-dynamics simulations~\cite{Spirit}, we test whether the exchange and DM interactions in B20 chiral magnets give rise to the twist of the spin-density waves, as described by Eq.~\eqref{model}. To prove the concept and to reduce the computational burden we consider first the magnetic interactions up to the fourth nearest neighbors, $|{\bf R}_{ij}^{AB}| \leq R_\text{max} = a$, where $a$ is the lattice parameter.  We use interaction parameters obtained for B20 FeGe and to reduce the size of the magnetic unit cell we multiply the DM interaction vectors by a factor of 10, see Fig.~\ref{fi:spirit}(a). 
We perform atomistic spin-dynamics simulations at zero temperature using cubic supercells of size $N_\lambda \times N_\lambda \times N_\lambda$, where $N_\lambda \in \{30, \ldots, 60\}$.  Our simulations confirm that the spin spirals form a twist at the ground state in all chosen supercells. Also, no modulations within the same sublattice in directions normal to the direction of ${\bf q}_{001}$ were observed; thus, a one-dimensional magnetic supercell $1 \times 1 \times N_\lambda$ is sufficient.

Figure~\ref{fi:spirit}(b) shows the total energies obtained from the minimization of the micromagnetic energy (solid lines), from the atomistic model (circles), and from the spin-dynamics simulations (diamonds) as a function of the period of the magnetic modulation $\lambda = 2\pi/\vert{\bf q}\vert$ for the spin spirals propagating in the $z$ direction.
Since the period of the magnetic modulations is rather small, we include micromagnetic parameters $\mi{M}{p}{AB}{}$ up to order $p=7$, which results in perfect agreement with an atomistic model. As shown in Fig.~\ref{fi:spirit}(b) the total energy minimum of the spin spiral is reduced by a factor of 14 due to the phase difference $\varphi$ (blue curve) and by a factor of 32 due to the deviation of the rotation axes  ${\bf e}^{A}_\text{rot}(\beta^A,\alpha^A)$ from the direction of ${\bf q}$ (red curve). The period of the magnetic modulations $\lambda_\text{min}$ becomes shorter due to such a twist as well. 
The corresponding twist angles $\varphi(\lambda)$, $\beta(\lambda)$, and $\alpha(\lambda)$ minimizing the micromagnetic total energy as a function of $\lambda$ are shown in Figs.~\ref{fi:spirit}(c).

Energies obtained by spin-dynamics simulations are only slightly lower than the energies predicted by our micromagnetic model, see Fig.~\ref{fi:spirit}(b).
This small difference is because in the micromagnetic model we assume homogeneous spirals for each sublattice, e.g., twist angles $\Omega^A =\{\phi^A,\beta^A, \alpha^A\}=\text{const}$ for any site ${\bf R}_i^A$.
However, spin-dynamics simulations show that $\Omega^A_i = \Omega^A + 
\Delta \Omega^A_i({\bf R}_{i}^{A})$, where $\Omega^A_i = f({\bf R}_i^A)$
is a function of the atomic position ${\bf R}_i^A$
\footnote{
The twist angles of the spirals in each sublattice $A$ are defined locally between the nearest magnetic sites ${\bf R}_i^A$ and ${\bf R}_{i-1}^A$ of the same sublattice:
$\phi^{A}_i = \arccos({\bf S}_i^A\cdot {\bf S}_{i-1}^A) - 
{\bf q}\cdot {\bf R}^{AA}_{i,i-1}$, and 
$\alpha^A_i$ and $\beta^A_i$ are polar coordinates 
of the rotation axis ${\bf e}_\text{rot}^A(\alpha^A_i,\beta^A_i) = {\bf S}^A_i\times {\bf S}^A_{i-1}$.}, see Fig.~\ref{fi:spirit}(d).
As can been seen from Fig.\ref{fi:spirit}, the energy gain due to such nonhomogeneity of the spin spirals $\Delta \Omega^A_i$ is much smaller than the energy gain due to a twist $\Omega^A$, as well as $\Delta \Omega^A_i\ll \Omega^A$.


On the other hand, the computational burden in the spin-dynamic simulations of the ground state properties of FeGe due to long range of the magnetic interactions can be significantly reduced by replacing the larger number of the realistic interaction parameters $\ro{J}{AB}{ij}$ by a smaller number of the  atomistic effective interaction parameters $\roti{J}{AB}{ij}$ with the condition that they both provide the same micromagnetic interaction tensors $\mi{M}{p}{AB}{}$ (see Appendix~\ref{sym}). Such  atomistic effective interaction parameters $\roti{J}{AB}{ij}$ can be obtained by solving the following system of equations with different orders $p$:
\begin{equation*}
\arraycolsep=1pt\def\arraystretch{1}
\begin{array}{lcl}
%
\frac{(-1)^{p+1}}{(2p)!} \disp 
\sum_{s=1}^{N_\text{sh}} \sum_{j=1}^{N_{B^s}} 
\roti{J}{AB^s}{1j(kk')}
\mi{R}{2p}{AB^s}{1j} 
&=& \mi{M}{2p}{AB}{kk'} \, ,
\\
%
\frac{(-1)^{p+1}}{(2p+1)!} \disp 
\sum_{s=1}^{N_\text{sh}} \sum_{j=1}^{N_{B^s}} 
\roti{J}{AB^s}{1j(kk')}
\mi{R}{2p+1}{AB^s}{1j}
&=&
\mi{M}{2p+1}{AB}{kk'}\, ,
\end{array}
\end{equation*}
where the sum over $s$ is restricted to  the number of the interacting (effective) shells $N_\text{sh}$ with radii $R^{s} = |{\bf R}_{ij}^{AB}| \leq R_\text{max}$ (shell index $s$ represent all bonds of the same length connected by symmetries) and $j$ runs over sites of sublattice $B^s$ within the same $R^s$.

The effective interaction parameters $\widetilde{J}_{ij}^{AB}$ and  $\widetilde{{\bf D}}_{ij}^{AB}$ for FeGe obtained as a fit to the micromagnetic tensors $\mi{A}{p}{AB}{}$ ($p\leq 2$) and $\mi{D}{p}{AB}{}$ ($p\leq 1$) are given in Table~\ref{eff}.  As shown in Fig.~\ref{fi:scale}~(a), the energies corresponding to these effective parameters obtained by SD simulations are in excellent agreement with the prediction of our micromagnetic model,  as well as with the energies obtained by the atomistic model applied to the predicted magnetic structure with realistic interaction parameters. However, while the prediction of our micromagnetic model is in excellent agreement with the SD simulations,  showing a significant reduction of the total energy of the spin spirals due to the twist, the obtained period of the magnetic modulations $\lambda_\text{min}^\text{DFT} \approx 140$ (nm) remains two times larger than the experimental observation.

A possible reason for this discrepancy might stem from the failure of DFT to capture the effect of electronic correlations, which can have a more drastic effect on DM interaction  than on the exchange, which provides  good agreement with the experimental Curie temperature~\cite{Grytsiuk2019}.
Therefore, we also study (using the micromagnetic model and SD simulations)  the dependences of the spin-spiral total energies, corresponding period, and twist angles on strength of the DM interaction vectors  $\eta {\bf D}_{ij}$, where $\eta = 0.9, \ldots 2.1$.
As shown in Fig.~\ref{fi:scale}, not only does the period $\lambda_\text{min}$ of the spin spiral become shorter when the strength of the DM interaction is increased, reaching the experimental $\lambda=70$ (nm) at $\eta \approx 2$, but also the corresponding twist angles $\phi_\text{min}$ and $\beta_\text{min}$ become larger, lowering the total energy $E_\text{min}$ more prominently than in the case when the twist of the spin spirals is ignored.

{\renewcommand{\arraystretch}{1.25}
\begin{table}[t!]\centering
\caption{
The effective interaction parameters $\widetilde{J}_{ij}^{AB^s}$ and  $\widetilde{{\bf D}}_{ij}^{AB^s}$  (where $s$ stands for the radius $R_s$ of the shell with all bonds 
$|{\bf R}_{ij}^{AB}| = R_s$ connected by symmetries) between atoms of sublattices $A=1$ and $B = 1, 2$ obtained as a fit to the micromagnetic parameters
$\mi{A}{p}{AB}{}$ ($p\leq 2$) and $\mi{D}{p}{AB}{}$ ($p\leq 1$). The effective interaction parameters between atoms of all other sublattices can be obtained by using symmetries (see Appendix~\ref{sym}).
}
\begin{tabular}{c|c|c|c|c|c}
\Xhline{1.5\arrayrulewidth}
$s$ &$A$-$B$ & $R_s$ & ${\bf R}_{ij}^{AB}$ & $\widetilde{J}_{ij}^{AB}$ & 
$\widetilde{{\bf D}}_{ij}^{AB}$
\\ \hline \hline
1 &1-2& 0.613 &   (0.23, -0.27, -0.5) &  3.311 & (-0.013, -0.295,  1.226)  \\
2 &1-2&  0.914 &  (0.23,  0.73, -0.5) &  1.474 &  (0.183,  0.088, -0.481)  \\
3 &1-2&  0.957 & (-0.77, -0.27, -0.5) & -0.006 &  (0.045, -0.175, -0.332)  \\
4 &1-2&  1.173 & (-0.77,  0.73, -0.5) & -0.263 &  --  \\
5 &1-2&  1.355 &  (1.23, -0.27, -0.5) &  1.997 &  --  \\
6 &1-2&  1.384 &  (0.23, -1.27, -0.5) & -0.101 &  --  \\
7 &1-2&  1.515 &  (1.23,  0.73, -0.5) & -1.421 &  --  \\
8 &1-2&  1.541 &  (0.23, -0.27, -1.5) &  1.694 &  --  \\\hline
1 &1-1&  1.000 & (-1,  0, 0) &  4.733 & (-0.127,  0.039,  0.108)  \\
2 &1-1&  1.414 & (-1, -1,  0) & -1.904 &  --  \\
3 &1-1&  1.414 & (-1,  0,  1) & -0.820 &  --  \\\Xhline{1.5\arrayrulewidth}
 \end{tabular}
\label{eff}
\end{table}}

\section{Magnetic structure}
\label{rotation}

We define the magnetic moment ${\bf S}^A_i$ at site $i$ of sublattice $A$ as
\begin{equation*}
\arraycolsep=1pt\def\arraystretch{1}
\begin{array}{lcl}
{\bf S}^A_i &=& 
\ro{R}{}{}(\beta^A,\alpha^A)
\begin{pmatrix} 
\sin \theta^A \cos [{\bf q}\cdot {\bf R}^A_i +\phi^A] \\
\sin \theta^A \sin [{\bf q}\cdot {\bf R}^A_i +\phi^A] \\
\cos \theta^A
 \end{pmatrix}\\
&=& 
 {\bf e}_1^A \cos ({\bf q}\cdot {\bf R}^A_i)
+ {\bf e}_2^A \sin ({\bf q}\cdot {\bf R}^A_i) +
{\bf e}_3^A\,,
\end{array}
\end{equation*}
where the parameter $\Omega^A = \{\beta^A,\alpha^A, \theta^A, \phi^A\}$ represents a set of spin-spiral twist angles and $\ro{R}{}{}(\beta^A,\alpha^A)$ is a rotation matrix mapping the ${\bf e}_z$ axis to the rotation axis ${\bf e}_\text{rot}^A$. Vectors ${\bf e}_m^A = {\bf e}_m^A(\Omega^A)$ characterizing the twist of the spin spiral in sublattice $A$ are defined as
\begin{equation*}
\begin{array}{lcl}
{\bf e}_1^A &=& 
\sin \theta^A\, \ro{R}{}{}(\beta^A,\alpha^A) 
\lra{
{\bf e}_x \cos \phi^A + {\bf e}_y \sin \phi^A}, \\
{\bf e}_2^A &=& 
\sin \theta^A\, \ro{R}{}{}(\beta^A,\alpha^A) 
\lra{
{\bf e}_y \cos \phi^A - {\bf e}_x \sin \phi^A}, \\
{\bf e}_3^A &=& 
\cos \theta^A\, \ro{R}{}{}(\beta^A,\alpha^A) \, {\bf e}_z = 
\cos \theta^A\, {\bf e}_\text{rot}^A .
\end{array}
\end{equation*}

\section{Trigonometric expansion}
\label{trig}

Since $({\bf q}\cdot {\bf R})^p = \mi{Q}{p}{}{}:\mi{R}{p}{}{}$, where $\mi{Q}{p}{}{}={\bf q}^{\bigotimes{p}}$ and $\mi{R}{p}{}{} = {\bf R}^{\bigotimes{p}}$ are the $p$-fold tensor products of ${\bf q}$ and ${\bf R}$ vectors with themselves, respectively, we expand the cosine  and sine functions of ${\bf q}\cdot {\bf R}$ in the form
\begin{equation*}
\arraycolsep=1pt\def\arraystretch{1}
 \begin{array}{lcl}
 -\cos {\bf q}\cdot {\bf R}&=& \disp 
 \sum_{p=0} \tfrac{(-1)^{p+1}}{(2p)!}\, \mi{Q}{2p}{}{} : \mi{R}{2p}{}{},\\
 -\sin {\bf q}\cdot {\bf R} &=& \disp 
 \sum_{p=0} \tfrac{(-1)^{p+1}}{(2p+1)!}\, \mi{Q}{2p+1}{}{} : \mi{R}{2p+1}{}{},
 \end{array}
\end{equation*}
where the sum over $p$ defines the order of Tailor expansion and and notation ``$:$" denotes the inner product between two tensors. Using this form for the expansion of the trigonometric functions 
the atomistic interaction tensors $\ro{J}{AB}{1j}$ entering the total energy expression (Eq.~\eqref{en3} in the main text) and corresponding distance between interacting sites 
${\bf R}^{AB}_{ij}$ can be combined in the micromagnetic interaction tensors $\mi{M}{p}{AB}{}$:
\begin{equation*}
\arraycolsep=1pt\def\arraystretch{1}
 \begin{array}{ll}
 &\disp
 -\sum_{j=1}\ro{J}{AB}{1j}\!:\! \ro{C}{AB}{+} 
 \cos {\bf q}\cdot {\bf R}_{1j}^{AB} \\
 &= \disp 
 \sum_{p=0} 
 \sum_{j=1}\tfrac{(-1)^{p+1}}{(2p)!}
 \lra{\ro{J}{AB}{1j}\!:\! \ro{C}{AB}{+}}
 \lra{\mi{Q}{2p}{}{}\!:\! \mi{R}{2p}{AB}{1j}}\\
 &=\disp 
 \sum_{p=0}
 \tfrac{(-1)^{p+1}}{(2p)!} \sum_{j=1}
 \lra{\ro{J}{AB}{1j}\otimes \mi{R}{2p}{AB}{1j}}
 \!:\!
 \lra{\ro{C}{AB}{+} \otimes \mi{Q}{2p}{}{}}\\
 &=\disp
 \sum_{p=0}
 \mi{M}{2p}{AB}{}\!:\! 
 \lra{\ro{C}{AB}{+} \otimes \mi{Q}{2p}{AB}{}}\, ,
 \end{array}
\end{equation*}
\begin{equation*}
\arraycolsep=1pt\def\arraystretch{1}
 \begin{array}{ll}
 &\disp
 -\sum_{j=1}\ro{J}{AB}{1j}\!:\! \ro{C}{AB}{-} 
 \sin {\bf q}\cdot {\bf R}_{1j}^{AB}\\ 
 &= \disp 
 \sum_{p=0} \sum_{j=1}\tfrac{(-1)^{p+1}}{(2p+1)!}
 \lra{\ro{J}{AB}{1j}\!:\!\ro{C}{AB}{-}}
 \lra{\mi{Q}{2p+1}{}{}\!:\!\mi{R}{2p+1}{AB}{1j}}
 \\
 & = \disp 
 \sum_{p=0}
 \tfrac{(-1)^{p+1}}{(2p+1)!} \sum_{j=1}
 \lra{\ro{J}{AB}{1j}\otimes 
 \mi{R}{2p+1}{AB}{1j}}
 \!:\! 
 \lra{\ro{C}{AB}{-}\!\otimes\!\mi{Q}{2p+1}{}{}}
 \\
 &=\disp
 \sum_{p=0} 
 \mi{M}{2p+1}{AB}{}\!:\! 
 \lra{\ro{C}{AB}{-}\!\otimes\!\mi{Q}{2p+1}{AB}{}}\, ,
 \end{array}
\end{equation*}
and
\begin{equation*}
\arraycolsep=1pt\def\arraystretch{1}
 \begin{array}{ll}
 &\disp
 -\sum_{j=1}\ro{J}{AB}{1j}\!:\!\ro{C}{AB}{} 
 =\disp
 \sum_{j=1}
 \lra{\ro{J}{AB}{1j}\!:\!\ro{C}{AB}{}}
 \lra{\mi{Q}{0}{}{}\!:\!\mi{R}{0}{AB}{1j}}\\
 &=\disp 
 \tfrac{-1}{0!} \sum_{j=1}
 \lra{\ro{J}{AB}{1j}\!\otimes\!\mi{R}{0}{AB}{1j}}
 \!:\!
 \lra{\ro{C}{AB}{}\!\otimes\! \mi{Q}{0}{}{}}\\
 &=
 \mi{M}{0}{AB}{}\!:\!\ro{C}{AB}{}\,,
 \end{array}
\end{equation*}
where tensors 
$\ro{C}{AB}{+} = {\bf e}_1^A \otimes {\bf e}_1^B + {\bf e}_2^A \otimes {\bf e}_2^B$, 
$\ro{C}{AB}{-} = {\bf e}_1^A \otimes {\bf e}_2^B - {\bf e}_2^A \otimes {\bf e}_1^B$, and
$\ro{C}{AB}{} = {\bf e}_3^A \otimes {\bf e}_3^B$ characterize the twist between spin spirals in sublattices $A$ and $B$.

\section{Symmetries and micromagnetic interaction tensors}
\label{sym}

{\renewcommand{\arraystretch}{1.25}
\begin{table*}[t!]\centering
\caption{
Symmetry relations between bonds 
${\bf R}_{i'j'}^{AB} = \ti{A}{\omega^{AB}} \cdot {\bf R}_{ij}^{12}$ and 
${\bf R}_{i'j'}^{AA'} = \ti{A}{\omega^{AB}} \cdot {\bf R}_{ij}^{11'}$ 
in the B20 structure.
}
\begin{tabular}{clc|clc|clc|clclclc}
\Xhline{1.5\arrayrulewidth}
$A$-$B$&$\ph{-}\ti{A}{0^\circ}$ & ${\bf R}^{AB}$ &
$A$-$B$&$\ph{-}\ti{A}{120^\circ}$ & ${\bf R}^{AB}$ &
$A$-$B$&$\ph{-}\ti{A}{240^\circ}$ & ${\bf R}^{AB}$ &
$A$-$A'$&$\ph{-}\ti{A}{0^\circ}$ & ${\bf R}^{AA'}$ 
 &$\ph{-}\ti{A}{120^\circ}$ & ${\bf R}^{AA'}$ 
 &$\ph{-}\ti{A}{240^\circ}$ & ${\bf R}^{AA'}$\\ \hline \hline
 \multirow{2}{*}{1-2}
 &$\ph{-}\ti{1}{0^\circ}$& ($x, y, z)$ &
 \multirow{2}{*}{1-3}
 &$\ph{-}\ti{1}{120^\circ}$& ($z, x, y)$ &
 \multirow{2}{*}{1-4}
 &$\ph{-}\ti{1}{240^\circ}$& ($y, z, x)$ & 
 \multirow{2}{*}{1-1}
 &$\ph{-}\ti{1}{0^\circ}$& ($x', y', z'$) 
 &$\ph{-}\ti{1}{120^\circ}$& ($z', x', y'$) 
 &$\ph{-}\ti{1}{240^\circ}$& ($y', z', x'$) \\
 &$-\ti{2}{0^\circ}$ & ($x, y, \bar{z})$ & 
 &$-\ti{3}{120^\circ}$ & ($\bar{z}, x, y)$ & 
 &$-\ti{4}{240^\circ}$ & ($y, \bar{z}, x)$ &
 &$-\ti{1}{0^\circ}$ & ($\bar{x}', \bar{y}', \bar{z}'$) 
 &$-\ti{1}{120^\circ}$ & ($\bar{z}', \bar{x}', \bar{y}'$) 
 &$-\ti{1}{240^\circ}$ & ($\bar{y}', \bar{z}', \bar{x}'$) \\ \hline
 \multirow{2}{*}{2-1}
 & $\ph{-}\ti{2}{0^\circ}$ &($\bar{x}, \bar{y}, z)$ &
 \multirow{2}{*}{2-4}
 & $\ph{-}\ti{2}{120^\circ}$ &($\bar{z}, \bar{x}, y)$ & 
 \multirow{2}{*}{2-3}
 & $\ph{-}\ti{2}{240^\circ}$ &($\bar{y}, \bar{z}, x)$ & 
 \multirow{2}{*}{2-2}
 & $\ph{-}\ti{2}{0^\circ}$&($\bar{x}', \bar{y}', z'$) 
 & $\ph{-}\ti{2}{120^\circ}$&($\bar{z}', \bar{x}', y'$) 
 & $\ph{-}\ti{2}{240^\circ}$&($\bar{y}', \bar{z}', x'$) \\
 &$-\ti{1}{0^\circ}$ & ($\bar{x}, \bar{y}, \bar{z})$ & 
 &$-\ti{4}{1}$ & ($z, \bar{x}, y)$ &
 &$-\ti{3}{240^\circ}$ & ($\bar{y}, z, x)$ & 
 &$-\ti{2}{0^\circ}$ & ($x', y', \bar{z}'$) 
 &$-\ti{2}{120^\circ}$ & ($z', x', \bar{y}'$) 
 &$-\ti{2}{240^\circ}$ & ($y', z', \bar{x}'$) \\ \hline
 \multirow{2}{*}{3-4} & $\ph{-}\ti{3}{0^\circ}$& ($x, \bar{y}, \bar{z})$ & 
 \multirow{2}{*}{3-1} & $\ph{-}\ti{3}{120^\circ}$& ($z, \bar{x}, \bar{y})$ &
 \multirow{2}{*}{3-2} & $\ph{-}\ti{3}{240^\circ}$& ($y, \bar{z}, \bar{x})$ &
 \multirow{2}{*}{3-3}
 &$\ph{-}\ti{3}{0^\circ}$ & ($x', \bar{y}', \bar{z}'$) 
 &$\ph{-}\ti{3}{120^\circ}$ & ($z', \bar{x}', \bar{y}'$) 
 &$\ph{-}\ti{4}{240^\circ}$ & ($y', \bar{z}', \bar{x}'$) \\
 &$-\ti{4}{0^\circ}$ & ($x, \bar{y}, z)$ &
 &$-\ti{1}{120^\circ}$ & ($\bar{z}, \bar{x}, \bar{y})$ & 
 &$-\ti{2}{240^\circ}$ & ($y, z, \bar{x})$ & 
 &$-\ti{3}{0^\circ}$ & ($\bar{x}', y', z'$) 
 &$-\ti{3}{120^\circ}$ & ($\bar{z}', x', y'$) 
 &$-\ti{3}{240^\circ}$ & ($\bar{y}', z', x'$) \\ \hline
 \multirow{2}{*}{4-3}
 &$\ph{-}\ti{4}{0^\circ}$ & ($\bar{x}, y, \bar{z})$ & 
 \multirow{2}{*}{4-2}
 &$\ph{-}\ti{4}{1}$ & ($\bar{z}, x, \bar{y})$ & 
 \multirow{2}{*}{4-1}
 &$\ph{-}\ti{4}{240^\circ}$ & ($\bar{y}, z, \bar{x})$ &
 \multirow{2}{*}{4-4}
 &$\ph{-}\ti{4}{0^\circ}$ & ($\bar{x}', y', \bar{z}'$) 
 &$\ph{-}\ti{4}{120^\circ}$ & ($\bar{z}', x', \bar{y}'$) 
 &$\ph{-}\ti{4}{240^\circ}$ & ($\bar{y}', z', \bar{x}'$) \\ 
 &$-\ti{3}{0^\circ}$ & ($\bar{x}, y, z)$ &
 &$-\ti{2}{120^\circ}$ & ($z, x, \bar{y})$ & 
 &$-\ti{1}{240^\circ}$ & ($\bar{y}, \bar{z}, \bar{x})$ & 
 &$-\ti{4}{0^\circ}$ & ($x', \bar{y}', z'$) 
 &$-\ti{4}{120^\circ}$ & ($z', \bar{x}', y'$) 
 &$-\ti{4}{240^\circ}$ & ($y', \bar{z}', x'$) \\ \hline
 \end{tabular}
\label{sym2}
\end{table*}}

\begin{figure}[t!]\center
 \includegraphics[width=0.48\textwidth]{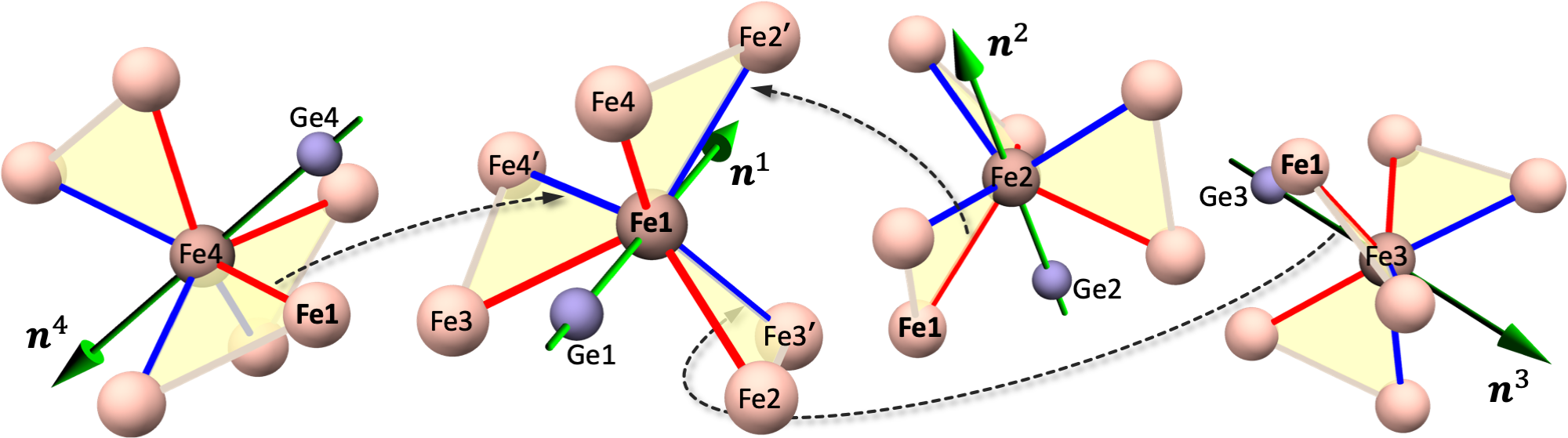}
 \caption{ 
 Nearest Fe (light red spheres) and Ge (blue spheres) atoms at each Fe site (darker red spheres) of the B20 structure.  Red and blue lines indicate bonds connected by threefold rotation symmetry with respect to the symmetry direction ${\bf n}^A$. 
 All red (blue) bonds ${\bf R}^{AB}_{ij}$ 
 (${\bf R}^{AB'}_{ij}$) are connected to 
 ${\bf R}^{12}_{ij}$ by $\ti{A}{\omega^{AB}}$ ($-\ti{B}{\omega^{AB}}$) symmetry.
 Dashed arrows indicate bonds at site $A=\text{Fe1}$ that appear at other sites 
 $B\neq A$.
 }
\label{NN}
\end{figure}

Magnetic and nonmagnetic atoms in B20 compounds are located at the 4$a$ Wyckoff positions, 
 ${\bf r}^1(u) =(u, u, u)$,
 ${\bf r}^2(u) =(0.5-u, 1-u, 0.5 + u)$,
 ${\bf r}^4(u) =(0.5 + u, 0.5-u, 1-u)$, and
 ${\bf r}^3(u) =(1-u, 0.5 + u, 0.5-u)$,
where the parameters $u^\text{Fe}=0.135$ and $u^\text{Ge}=-0.158$ stand, respectively, for Fe and Ge atoms in the B20 FeGe compound. Each sublattice $A$ in B20 compounds has only one threefold rotation axis (see Fig.~\ref{NN}), which can be defined as
\begin{equation*}
 {\bf n}^{A} = \frac{\partial {\bf r}^A(u)}{\partial u}
 \in \{(111), (\bar{1}\bar{1}1), (1\bar{1}\bar{1}), (\bar{1}1\bar{1})\}\,.
\end{equation*}
All symmetries at any site of any sublattice $A$ in the B20 structure can be generalized  by the following symmetry operation
\begin{equation}\label{rotsym}
\ti{AB}{} = 
\ti{A}{\omega^{AB}} = \lra{{\bf n}^A\otimes {\bf n}^A}\circ \ro{R}{[111]}{\omega^{AB}},
\end{equation}
where $\ro{R}{[111]}{\omega^{AB}}$ is a threefold rotation matrix around the ${\bf n}^1=[1,1,1]$ axis on angle $\omega^{AB} = \frac{2\pi}{3}k$ ($k = 0, 1, 2$), ``$\circ$" is the elementwise (Hadamard) product, and operation $\lra{{\bf n}^A\otimes {\bf n}^A}$ maps sublattice $1$ to any other sublattice $A$,
such that any bond ${\bf R}^{AB}_{i'j'}$ between sites $i'$ and $j'$ of any sublattices $A$ and $B$, respectively, can be represented as ${\bf R}^{AB}_{i'j'} = \ti{AB}{} \cdot {\bf R}^{12}_{ij}$ or as ${\bf R}^{AA}_{i'j'} = \ti{AA}{} \cdot {\bf R}^{11}_{ij}$ if $A=B$.
Note that the successive three fold rotations around ${\bf n}^A$ gives ${\bf R}_{ij}^{AB} \rightarrow {\bf R}_{ik}^{AC} \rightarrow {\bf R}_{il}^{AD}$,  where the order of  sublattices $B$, $C$, $D$ is such that ${\bf n}^A\cdot ({\bf R}_{ij}^{AB} \times {\bf R}_{ik}^{AC})\geq0$ and ${\bf n}^A\cdot ({\bf R}_{ik}^{AC} \times {\bf R}_{il}^{AD})\geq0$.
Also, it is easy to show that  the operation $-\ti{B}{\omega^{AB}}$ at sublattice $A$ gives ${\bf R}^{AB}_{i'j'}= -\ti{B}{\omega^{AB}}{\bf R}^{12}_{ij}= -{\bf R}^{BA}_{j'i'}$. It represents double counting
and therefore can be omitted;  otherwise, a factor of $1/2$ has to be used in the energy calculations. For instance, any site of sublattice $A$ has six nearest neighbors in total; however, only three of them are related by threefold rotation symmetry ${\bf n}^A$, and the other three appear at sites of other sublattices $B$, $C$, and $D$ (see Fig.~\ref{NN}). 
The symmetry operations  $\ti{A}{\omega^{AB}}$ and $-\ti{B}{\omega^{AB}}$ for the bonds are illustrated in Table~\ref{sym2}, and those for the micromagnetic vectors $\mi{D}{0}{AB}{}$ and $\mi{A}{1}{AB}{} $ are given in Table~\ref{sym3}.

{\renewcommand{\arraystretch}{1.25}
\begin{table*}[t!]\centering
\caption{Micromagnetic tensors 
$\mi{D}{0}{AB}{}$ and 
$\mi{A}{1}{AB}{}$ between different pairs of sublattices $A$ and $B$ of B20 FeGe.}
\begin{tabular}{c|cr|cr|cr|cr}
\Xhline{1.5\arrayrulewidth}
 \mi{M}{p}{AB}{} 
 & A-B& $(x,y,z)$ 
 & A-B& $(x,y,z)$ 
 & A-B& $(x,y,z)$ 
 & A-A& $(x,y,z)$ \\ \hline \hline
%
%
 \multirow{4}{*}{$\mi{D}{0}{AB}{}$ (meV)
 }
 & {1-2} & $(-0.43,0.76,0)$ 
 & {1-3} & $(0,-0.43,0.76)$ 
 & {1-4} & $(0.76,0,-0.43)$ 
 & {1-1} & $(0, 0, 0$) \\ 
%
%
 & {2-1} & $(0.43, -0.76, 0)$ 
 & {2-4} & $(0, 0.43, 0.76)$ 
 & {2-3} & $(-0.76, 0, -0.43)$ 
 & {2-2} & $(0, 0, 0$) \\ 
%
%
 & {3-4} & $(-0.43, -0.76, 0)$ 
 & {3-1} & $(0, 0.43, -0.76)$ 
 & {3-2} & $(0.76, 0, 0.43)$ 
 & {3-3} & $(0, 0, 0$) \\ 
%
%
 & {4-3} & $(0.43, 0.76, 0)$ 
 & {4-2} & $(0, -0.43, -0.76)$ 
 & {4-1} & $(-0.76, 0, 0.43)$ 
 & {4-4} & $(0, 0, 0$) \\ \hline
%
%
%
\multirow{4}{*}{$\mi{A}{1}{AB}{}$ (meV\AA)}
&1-2 & $(-22.39, 17.99, 0)$ 
&1-3 & $(0, -22.39, 17.99)$ 
&1-4 & $(17.99, 0, -22.39)$ 
&1-1 & $(0, 0, 0)$ \\ 
%
%
&2-1 & $(22.39, -17.99, 0)$ 
&2-4 & $(0, 22.39, 17.99)$ 
&2-3 & $(-17.99, 0, -22.39)$ 
&2-2 & $(0, 0, 0)$ \\ 
%
%
&3-4 & $(-22.39, -17.99, 0)$ 
&3-1 & $(0, 22.39, -17.99)$ 
&3-2 & $(17.99, 0, 22.39)$ 
&3-3 & $(0, 0, 0)$ \\ 
%
%
&4-3 & $(22.39, 17.99, 0)$ 
&4-2 & $(0, -22.39, -17.99)$ 
&4-1 & $(-17.99, 0, 22.39)$ 
&4-4 & $(0, 0, 0)$ \\ \hline
\end{tabular}
\label{sym3}
\end{table*}}

Since symmetry-related bonds ${\bf R}^{AB}_{i'j'}$ and symmetry-related DM vectors ${\bf D}^{AB}_{i'j'}$ can be defined via  ${\bf R}^{12}_{ij}$ or ${\bf R}^{11}_{ij}$ and ${\bf D}^{12}_{ij}$ or ${\bf D}^{11}_{ij}$, respectively,  all micromagnetic tensors $\mi{M}{p}{AB}{}$ can be also defined in terms of $\mi{M}{p}{12}{}$ or $\mi{M}{p}{11}{}$:
\begin{equation}\label{misym1}
\arraycolsep=1pt\def\arraystretch{1}
\left\{
\begin{array}{lcl}
\mi{A}{0}{AB}{}
&=& \mi{A}{0}{12}{},\quad
\mi{A}{1}{AB}{} = \ti{AB}{}\mi{A}{1}{12}{}, \\
\mi{A}{2}{AB}{} &=& \ti{AB}{} \mi{A}{2}{12}{} (\ti{AB}{})^T,\\
\mi{A}{3}{AB}{} &=&
\ti{AB}{}
\Big(
\ti{AB}{}
\big(
\mi{A}{3}{12}{}
\big)^T\Big)^T
\big(\ti{AB}{}\big)^{T}.
\end{array}\right.\\
\end{equation}
The same operations are valid for tensors $\mi{D}{p}{}{}$. Note if $A=B$  the only symmetry operation $\ti{A}{0}$ is required to transform  $\mi{M}{0}{11}{} $ to $\mi{M}{0}{AA}{}$. Below we list tensors $\mi{M}{p}{12}{}$ and $\mi{M}{p}{11}{}$ for B20 FeGe as well as the sum of $\mi{M}{p}{AB}{}$ over all pairs of sublattices:

\begin{equation*}
\arraycolsep=1.pt\def\arraystretch{1.25}
\footnotesize
\begin{tabular}{c|c|c|c}\Xhline{1.5\arrayrulewidth}
$AB$ & $\mi{A}{0}{AB}{}$ & $\mi{A}{1}{AB}{}$ & $\mi{D}{0}{AB}{}$\\\hline
$11$ & $-12.06$ & $(0,0,0)$& $(0,0,0)$\\\hline
$12$ & $-13.37$ & $(-22.39,17.99,0)$ & $(-0.43,0.76,0)$\\\hline
$\sum_{AB}$ & $-208.68$ & $(0,0,0)$& $(0,0,0)$\\
\end{tabular}
\end{equation*}

\begin{equation*}
\arraycolsep=1.pt\def\arraystretch{1.25}
\footnotesize
\begin{tabular}{c|c|c}\Xhline{1.5\arrayrulewidth}
$AB$ & $\mi{A}{2}{AB}{}$ & $\mi{D}{1}{AB}{}$\\\hline
$11$ & 
$\ta{1ex}{r}{
-15.77 & -23.96 & -23.96 \\
-23.96 & -15.77 & -23.96 \\
-23.96 & -23.96 & -15.77}$
&
$\ta{1ex}{r}{
-1.20 & 1.01 & 0.37 \\
 0.37 & -1.20 & 1.01 \\
 1.01 & 0.37 & -1.20}$\\\hline
$12$ &
$\ta{1ex}{r}{
 23.15 &-40.36 & \ph{-1}0.00 \\
-40.36 & 5.19 & \ph{-1}0.00 \\
 0.00 & 0.00 & 111.78 } $
&
$\ta{1ex}{r}{
-0.04 &-1.17 & \ph{-}0.00\\
-0.82 &-1.80 & 0.00\\
 0.00 & 0.00 & 1.94}$ \\\hline
$\sum_{AB}$ & $497.42\ro{I}{}{1}{}$& $-4.38\ro{I}{}{1}{}$
\end{tabular}
\end{equation*}

\begin{equation*}
\arraycolsep=1.pt\def\arraystretch{1.25}
\footnotesize
\begin{tabular}{c|c|c|c}
\Xhline{1.5\arrayrulewidth}
\multicolumn{2}{c|}{$\mi{A}{3}{12}{}$} & \multicolumn{2}{c}{$\mi{D}{2}{12}{}$} \\\hline
\multicolumn{2}{c|}{$\begin{pmatrix}
\ta{1ex}{r}{
-305.74 & -14.10 & \ph{-1}0.00\\
 -14.10 & \ph{-1}81.11 & \ph{-1}0.00\\
\ph{-}0.00 & \ph{-}0.00 & \ph{-1}77.91
}\\
\ta{1ex}{r}{
-14.10 & \ph{-}81.11 & 0.00\\
\ph{-1}81.11 & \ph{-}345.57 & 0.00\\
0.00 & 0.00 & -140.59
}\\
\ta{1ex}{r}{
0.00 & 0.00 & 77.91 \\
0.00 & 0.00 & -140.59 \\
\ph{-1}77.91 &-140.59 & 0.00
}
\end{pmatrix}$}
&
\multicolumn{2}{c}{$\begin{pmatrix}
\ta{1ex}{r}{
17.27 & \ph{-1}0.07 & \ph{-}0.00\\
\ph{-}0.07 & -4.89 & 0.00\\
\ph{-}0.00 & 0.00 & -0.93
}\\
\ta{1ex}{r}{
5.42 & -2.56 & \ph{-}0.00\\
-2.56 & -18.61 & 0.00\\
\ph{-}0.00 & 0.00 & 0.17 
}\\
\ta{1ex}{r}{
 0.00 & \ph{-}0.00 & -4.72\\
 \ph{-}0.00 & \ph{-1}0.00 & 6.54\\
 -4.72 & \ph{-}6.54 & 0.00
}
\end{pmatrix}$}\\\hline
\multicolumn{2}{c|}{$\mi{A}{3}{11}{}=
\sum_{AB}\mi{A}{3}{AB}{}=0\ro{I}{}{3}{}$}& 
\multicolumn{2}{c}{$\mi{D}{2}{11}{}=
\sum_{AB}\mi{D}{2}{AB}{}=0\ro{I}{}{3}{}$}\\
\end{tabular}
\end{equation*}
Here, $\mathcal{I}_p$ is identity tensor of rank $p$. As we can see, the sum of the above tensors $\mi{M}{p}{AB}{}$ over all pairs of sublattices of B20 structure gives the scalar matrices due to the symmetry of the B20 structure characterized by the point group $T$. Similarly, summing up the higher-order tensors $\mi{D}{3}{AB}{}$ and $\mi{A}{4}{AB}{}$ over all pairs of sublattices gives
\begin{widetext}
\begin{equation}\label{At3}
\begin{array}{lcl}
\mi{D}{3}{}{} &=& 
\disp \sum_{AB} \mi{D}{3}{AB}{} = 
%
\arraycolsep=0.5pt\def\arraystretch{0.5}{
\begin{pmatrix}
 \setstacktabbedgap{1ex}
 \parenMatrixstack[c]{
 D^{(3)}_1 & \ph{I_{(1)}}0\ph{I} & \ph{I_{(1)}}0\ph{I} \\
 \ph{I_{(1)}}0\ph{I} & D^{(3)}_2 & \ph{I_{(1)}}0\ph{I} \\
 \ph{I_{(1)}}0\ph{I} & \ph{I_{(1)}}0\ph{I} & D^{(3)}_3}
 \setstacktabbedgap{1ex}
 \parenMatrixstack[c]{
 \ph{I_{(1)}}0\ph{I} & D^{(3)}_2 & \ph{I_{(1)}}0\ph{I} \\
 D^{(3)}_2 & \ph{I_{(1)}}0\ph{I} & \ph{I_{(1)}}0\ph{I} \\
 \ph{I_{(1)}}0\ph{I} & \ph{I_{(1)}}0\ph{I} & \ph{I_{(1)}}0\ph{I}}
 \setstacktabbedgap{1ex}
 \parenMatrixstack[c]{
 \ph{I_{(1)}}0\ph{I} & \ph{I_{(1)}}0\ph{I} & D^{(3)}_3 \\
 \ph{I_{(1)}}0\ph{I} & \ph{I_{(1)}}0\ph{I} & \ph{I_{(1)}}0\ph{I} \\
 D^{(3)}_3 & \ph{I_{(1)}}0\ph{I} & \ph{I_{(1)}}0\ph{I}
 }\\
 \setstacktabbedgap{1ex}
 \parenMatrixstack[c]{
 \ph{I_{(1)}}0\ph{I} & D^{(3)}_3 & \ph{I_{(1)}}0\ph{I} \\
 D^{(3)}_3 & \ph{I_{(1)}}0\ph{I} & \ph{I_{(1)}}0\ph{I} \\
 \ph{I_{(1)}}0\ph{I} & \ph{I_{(1)}}0\ph{I} & \ph{I_{(1)}}0\ph{I}}
 \setstacktabbedgap{1ex}
 \parenMatrixstack[c]{
 D^{(3)}_3 & \ph{I_{(1)}}0\ph{I} & \ph{I_{(1)}}0\ph{I} \\
 \ph{I_{(1)}}0\ph{I} & D^{(3)}_1 & \ph{I_{(1)}}0\ph{I} \\
 \ph{I_{(1)}}0\ph{I} & \ph{I_{(1)}}0\ph{I} & D^{(3)}_2}
 \setstacktabbedgap{1ex}
 \parenMatrixstack[c]{
 \ph{I_{(1)}}0\ph{I} & \ph{I_{(1)}}0\ph{I} & \ph{I_{(1)}}0\ph{I} \\
 \ph{I_{(1)}}0\ph{I} & \ph{I_{(1)}}0\ph{I} & D^{(3)}_2 \\
 \ph{I_{(1)}}0\ph{I} & D^{(3)}_2 & \ph{I_{(1)}}0\ph{I}
 }\\
 \setstacktabbedgap{1ex}
 \parenMatrixstack[c]{
 \ph{I_{(1)}}0\ph{I} & \ph{I_{(1)}}0\ph{I} & D^{(3)}_2 \\
 \ph{I_{(1)}}0\ph{I} & \ph{I_{(1)}}0\ph{I} & \ph{I_{(1)}}0\ph{I} \\
 D^{(3)}_2 & \ph{I_{(1)}}0\ph{I} & \ph{I_{(1)}}0\ph{I}}
 \setstacktabbedgap{1ex}
 \parenMatrixstack[c]{
 \ph{I_{(1)}}0\ph{I} & \ph{I_{(1)}}0\ph{I} & \ph{I_{(1)}}0\ph{I} \\
 \ph{I_{(1)}}0\ph{I} & \ph{I_{(1)}}0\ph{I} & D^{(3)}_3 \\
 \ph{I_{(1)}}0\ph{I} & D^{(3)}_3 & \ph{I_{(1)}}0\ph{I}}
 \setstacktabbedgap{1ex}
 \parenMatrixstack[c]{
 D^{(3)}_2 & \ph{I_{(1)}}0\ph{I} & \ph{I_{(1)}}0\ph{I} \\
 \ph{I_{(1)}}0\ph{I} & D^{(3)}_3 & \ph{I_{(1)}}0\ph{I} \\
 \ph{I_{(1)}}0\ph{I} & \ph{I_{(1)}}0\ph{I} & D^{(3)}_1}\\
\end{pmatrix}},\\
&&\\
\mi{A}{4}{}{} &=& 
\disp \sum_{AB} \mi{A}{4}{AB}{} = 
\arraycolsep=1pt\def\arraystretch{1}{
\begin{pmatrix}
 \setstacktabbedgap{1ex}
 \parenMatrixstack[c]{
 A^{(4)}_1 & \ph{I_{(1)}}0\ph{I} & \ph{I_{(1)}}0\ph{I} \\
 \ph{I_{(1)}}0\ph{I} & A^{(4)}_2 & \ph{I_{(1)}}0\ph{I} \\
 \ph{I_{(1)}}0\ph{I} & \ph{I_{(1)}}0\ph{I} & A^{(4)}_2}
 \setstacktabbedgap{1ex}
 \parenMatrixstack[c]{
 \ph{I_{(1)}}0\ph{I} & A^{(4)}_2 & \ph{I_{(1)}}0\ph{I} \\
 A^{(4)}_2 & \ph{I_{(1)}}0\ph{I} & \ph{I_{(1)}}0\ph{I} \\
 \ph{I_{(1)}}0\ph{I} & \ph{I_{(1)}}0\ph{I} & \ph{I_{(1)}}0\ph{I}}
 \setstacktabbedgap{1ex}
 \parenMatrixstack[c]{
 \ph{I_{(1)}}0\ph{I} & \ph{I_{(1)}}0\ph{I} & A^{(4)}_2 \\
 \ph{I_{(1)}}0\ph{I} & \ph{I_{(1)}}0\ph{I} & \ph{I_{(1)}}0\ph{I} \\
 A^{(4)}_2 & \ph{I_{(1)}}0\ph{I} & \ph{I_{(1)}}0\ph{I}
 }\\
 \setstacktabbedgap{1ex}
 \parenMatrixstack[c]{
 \ph{I_{(1)}}0\ph{I} & A^{(4)}_2 & \ph{I_{(1)}}0\ph{I} \\
 A^{(4)}_2 & \ph{I_{(1)}}0\ph{I} & \ph{I_{(1)}}0\ph{I} \\
 \ph{I_{(1)}}0\ph{I} & \ph{I_{(1)}}0\ph{I} & \ph{I_{(1)}}0\ph{I}}
 \setstacktabbedgap{1ex}
 \parenMatrixstack[c]{
 A^{(4)}_2 & \ph{I_{(1)}}0\ph{I} & \ph{I_{(1)}}0\ph{I} \\
 \ph{I_{(1)}}0\ph{I} & A^{(4)}_1 & \ph{I_{(1)}}0\ph{I} \\
 \ph{I_{(1)}}0\ph{I} & \ph{I_{(1)}}0\ph{I} & A^{(4)}_2}
 \setstacktabbedgap{1ex}
 \parenMatrixstack[c]{
 \ph{I_{(1)}}0\ph{I} & \ph{I_{(1)}}0\ph{I} & \ph{I_{(1)}}0\ph{I} \\
 \ph{I_{(1)}}0\ph{I} & \ph{I_{(1)}}0\ph{I} & A^{(4)}_2 \\
 \ph{I_{(1)}}0\ph{I} & A^{(4)}_2 & \ph{I_{(1)}}0\ph{I}
 }\\
 \setstacktabbedgap{1ex}
 \parenMatrixstack[c]{
 \ph{I_{(1)}}0\ph{I} & \ph{I_{(1)}}0\ph{I} & A^{(4)}_2 \\
 \ph{I_{(1)}}0\ph{I} & \ph{I_{(1)}}0\ph{I} & \ph{I_{(1)}}0\ph{I} \\
 A^{(4)}_2 & \ph{I_{(1)}}0\ph{I} & \ph{I_{(1)}}0\ph{I}}
 \setstacktabbedgap{1ex}
 \parenMatrixstack[c]{
 \ph{I_{(1)}}0\ph{I} & \ph{I_{(1)}}0\ph{I} & \ph{I_{(1)}}0\ph{I} \\
 \ph{I_{(1)}}0\ph{I} & \ph{I_{(1)}}0\ph{I} & A^{(4)}_2 \\
 \ph{I_{(1)}}0\ph{I} & A^{(4)}_2 & \ph{I_{(1)}}0\ph{I}}
 \setstacktabbedgap{1ex}
 \parenMatrixstack[c]{
 A^{(4)}_2 & \ph{I_{(1)}}0\ph{I} & \ph{I_{(1)}}0\ph{I} \\
 \ph{I_{(1)}}0\ph{I} & A^{(4)}_2 & \ph{I_{(1)}}0\ph{I} \\
 \ph{I_{(1)}}0\ph{I} & \ph{I_{(1)}}0\ph{I} & A^{(4)}_1}\\
\end{pmatrix}}
,\\
\end{array}
\end{equation}
where 
$D^{(3)}_1 = -140.86\, \text{meV\AA}^3$,
$D^{(3)}_2 = 27.16\, \text{meV\AA}^3$,
$D^{(3)}_3 = 27.92\, \text{meV\AA}^3$,
$A^{(4)}_1 = -1195.94\, \text{meV\AA}^4$,
and
$A^{(4)}_2 = -126.49\, \text{meV\AA}^4$.
\end{widetext}

\onecolumngrid

\section{Micromagnetic energies}
\label{energies}

The total energy of the spin spirals with twist, Eq.~\eqref{en4}, can be unfolded into the exchange and DM contributions:
\begin{equation*}\label{unfold1}
\arraycolsep=1pt\def\arraystretch{1}{
 \begin{array}{lcl}
 E^{AB}_\text{ex}(\bf q) &=& \disp
 \mi{A}{0}{AB}{} 
 \lra{{\bf e}_3^A \cdot {\bf e}_3^B}
 +\disp
 \frac{1}{2} \disp \sum_{mn}^2 
 \Big({\bf e}_m^A \cdot {\bf e}_n^B\Big) 
 \Big[
 \delta_{mn} \mi{A}{0}{AB}{} 
 +
 \varepsilon_{mn} \mi{A}{1}{AB}{} \cdot {\bf q}
 +
 \delta_{mn} \mi{A}{2}{AB}{} : {\bf q}^{\otimes 2}
 + \cdots
 \Big]
 \\
 E^{AB}_\text{dm}(\bf q) &=& \disp
 \mi{D}{0}{AB}{} \cdot
 \lra{{\bf e}_3^A \times {\bf e}_3^B}
 +\disp
 \frac{1}{2} \disp \sum_k \sum_{mn}^2 
 \Big({\bf e}_m^A \times {\bf e}_n^B\Big)_k 
 \Big[
 \delta_{mn} \mi{D}{0}{AB}{k} 
 +
 \varepsilon_{mn} \mi{D}{1}{AB}{k} \cdot {\bf q}
 +
 \delta_{mn} \mi{D}{2}{AB}{k} : {\bf q}^{\otimes 2}
 + \cdots
 \Big]
 \end{array}}
\end{equation*}
where indexes $k=x,y,z$ represent the components of the micromagnetic DM tensor $\mi{D}{p}{AB}{k}$ and symbols $\varepsilon_{mn}$ and $\delta_{mn}$ are the antisymmetric two-dimensional Levi-Civita tensor and Kronecker delta function, respectively, for which each index $m,n$ takes values $1,2$.

\twocolumngrid

\bibliography{references}

\end{document}